\documentclass{jfm}
\usepackage{soul}
\usepackage{graphicx}
\usepackage{newtxtext}
\usepackage{newtxmath}
\usepackage{amsmath} %
\usepackage{bm} %
\usepackage{natbib}
\usepackage{hyperref}
\usepackage{booktabs}
\usepackage{soul, color, xcolor}
\hypersetup{
	colorlinks = true,
	urlcolor = blue,
	citecolor = black,
}

\newcommand{\RomanNumeralCaps}[1]

\soulregister{\cite}7 % 注册\cite命令

\usepackage{soul}
\soulregister\cite7
\soulregister\ref7 
\soulregister\citep7 % 针对\citep命令
\soulregister\citet7 % 针对\citet命令
\soulregister\pageref7 % 针对\pageref命令

\title{On the role of inertia and self-sustaining mechanism in two-dimensional elasto-inertial turbulence}
\author{Haotian Cheng\aff{1}, Hongna Zhang\aff{1,2}\corresp{\email{hongna@tju.edu.cn}} , Wenhua Zhang\aff{2}\corresp{\email{zhangwh2022@tju.edu.cn}}, Yuke Li\aff{3}, Xiaobin Li\aff{1} \and Fengchen Li\aff{1}}

\affiliation{\aff{1}State Key Laboratory of Engine, Tianjin University, Tianjin 300350, PR China \aff{2}School of Petroleum Engineering, Yangtze University, Wuhan 430100, Hubei, PR China \aff{3}College of Shipbuilding Engineering, Harbin Engineering University, Harbin 150001, PR China and National Key Laboratory of Ship Structural Safety, Harbin 150001, Heilongjiang, PR China}
\begin{document}
	\maketitle
	
	\begin{abstract}
	Elasto-inertial turbulence (EIT) is primarily driven by polymer elasticity, yet the modulating role of fluid inertia is non-negligible and remains largely unexplored. To investigate the effect of inertia, we perform direct numerical simulations of two-dimensional EIT in channel flow over a wide range of Reynolds numbers ($Re$). We show that increasing inertia promotes both the enhancement of dynamic amplitudes and the wallward migration of core structures. Specifically, inertia intensifies the turbulent fluctuations, facilitates the fragmentation of large-scale structures, and amplifies statistical quantities such as the root-mean-square of velocity fluctuations and polymer extension. The peak location of nonlinear elastic shear stress follows a scaling law $y^+ \propto Re_\tau^{1/2}$, closely resembling that of Reynolds shear stress in Newtonian turbulence, indicating a change of the momentum transfer mechanism. Meanwhile, the peak location of energy conversion between elastic and turbulent kinetic energies exhibits a $y^+ \propto Re_\tau^{0.1}$ scaling law migration, remaining mostly confined to the near-wall region. Remarkably, despite the inertial modulation, the probability density functions (PDFs) of velocity and elastic stress fluctuations extracted at the energy-conversion peak collapse convincingly over the range of $Re$ investigated. This reveals a robust statistical self-similarity across a wide range of inertia magnitude. Furthermore, the PDFs of wall-normal velocity and elastic stress fluctuations exhibit pronounced exponential heavy tails. Combined with instantaneous flow topologies, these extreme events elucidate a universal self-sustaining mechanism for EIT: forward sweep events (Q1 motions, positive wall-normal velocity fluctuations) induce strong polymer stretching and storage of elastic energy, while local backward impacts (Q3 events, negative wall-normal velocity fluctuations) trigger the relaxation and rupture of polymer extension sheets, releasing elastic energy to regenerate turbulent kinetic energy.
	\end{abstract}
	
	%{\bf MSC Codes } {\it(Optional)} Please enter your MSC Codes here	

\section{Introduction}

Viscoelastic fluids are ubiquitous in nature and modern industrial applications (Bird et al. 1987; White and Mungal 2008). Initially, Toms (1948) discovered that the addition of polymers into Newtonian fluids to form viscoelastic fluids can achieve turbulent drag reduction; as the concentration further increases, the drag reduction rate approaches the maximum drag reduction (MDR) asymptote (Virk et al. 1970). For over half a century, the physical nature of MDR and its underlying dynamical mechanisms have remained recognized core challenges in the fluid mechanics community (Lumley 1969; Graham 2014; Xi 2019). The introduction of elasticity can fundamentally alter flow stability and turbulent states. Researchers discovered `elastic turbulence' (ET), induced by purely elastic nonlinear instabilities, in the extreme parameter space of very low Reynolds number ($Re\to0$) and high Weissenberg number ($Wi$) (Groisman and Steinberg 2000; Steinberg 2021). Furthermore, considering the coupling effect of elasticity and inertia, Samanta et al. (2013) and Dubief et al. (2013) proposed a new turbulent state named elasto-inertial turbulence (EIT) at moderate Re. EIT highly coincides with the MDR state in terms of structural and statistical properties, so it is widely regarded as the key flow state to unravel the physical essence of MDR. However, precisely because EIT is closely related to the highly elasticity-dominated MDR and the purely elasticity-driven ET, EIT-related studies have long focused their attention on the role of polymer elasticity, largely neglecting the specific role that inertia plays in EIT.

Since ET completely excludes inertial nonlinearity, the exploration of its underlying mechanisms naturally focuses on polymer elasticity (Burghelea et al. 2006; Steinberg 2021). However, for EIT, which inherently incorporates inertial effects by definition, the research focus still leans heavily towards the elasticity aspect. Regarding the self-sustaining mechanism, Dubief et al. (2013) first revealed that elastic stress in EIT can do reverse work to inject energy into the turbulence, thereby sustaining turbulence at subcritical $Re$. Subsequently, Terrapon et al. (2015) found through pressure decomposition that polymer pressure facilitates the maintenance of the quasi-two-dimensional characteristics of EIT. Sid et al. (2018) confirmed that EIT can self-sustain in two-dimensional (2D) DNS, making 2D numerical simulations, which filter out inertial turbulence, a reasonable approach for studying EIT (Zhu and Xi 2021; Zhang et al. 2024). ET can similarly self-sustain in 2D DNS (Berti et al. 2008). Zhang et al. (2021) systematically revealed the self-sustaining energy cycle of EIT: the average kinetic energy is first converted into streamwise elastic energy via nonlinear elastic shear stress, followed by energy conversion through polymer-turbulence interactions, ultimately feeding back to generate nonlinear elastic shear stress. Regarding the instability mechanism, early studies debated between two types of instabilities: Garg et al. (2018) and Page et al. (2020) discovered the center-mode instability triggered by nonlinear elasto-inertial traveling waves, while Shekar et al. (2019, 2021) identified the wall-mode instability based on Tollmien--Schlichting (T-S) waves. However, recent nonlinear dynamics studies point out that coherent structures with elastic nature, such as arrowhead structures, may act as attractors in the flow field, and the true trigger origin of EIT still points to the wall mode (Beneitez et al. 2024; Couchman et al. 2024; Zhang et al. 2024). Furthermore, regarding the nature of MDR, EIT is almost identical to the MDR state in terms of structural and statistical properties (Dubief et al. 2013), and experiments have proven the objective existence of EIT (Choueiri et al. 2018). It is generally believed that MDR is not a remnant state of Newtonian turbulence, and is even equivalent to EIT (Sid et al. 2018). However, Zhu and Xi (2021) proposed that MDR is a comprehensive state containing various dynamical patterns partially sustained by polymer elasticity, with EIT being just one of them; in our recent work, we also argued that MDR is not equivalent to EIT through the phenomenon of anomalous Reynolds stress (Cheng et al. 2025). Although debates and discussions continue, the aforementioned studies all indicate that polymer elasticity is the dominant dynamical factor driving EIT.

After confirming the dominant role of elasticity, the intrinsic connection between ET and EIT, two similar flows, has become the frontier of current research. Recent linear stability analyses have found the center-mode instabilities in both ET and EIT (Garg et al. 2018; Khalid et al. 2021), revealing a theoretical connection between the two at the linear theory level. Regarding the transition mechanisms, Beneitez et al. (2024b) recently revealed a common Ruelle-Takens transition scenario for both flow states in 2D channel flow, demonstrating that wall-localized traveling waves can trigger large-scale secondary instabilities to sustain the chaotic dynamics. At the level of nonlinear coherent structures, similar arrowhead structures (Dubief et al. 2022) and ``narwhal" traveling-wave solutions (Lellep et al., 2024) have been captured in both 2D and3D DNS, which are regarded as crucial physical structures connecting ET and EIT. More notably, recent experimental and numerical evidence demonstrates a high degree of structural consistency and spatial evolution characteristics between ET and EIT. Rosti et al. (2024) found in 3D channel DNS that as $Re$ decreases from 2800 to 0.5, once inertia recedes to a secondary role ($Re \le 1500$), the turbulent kinetic energy spectra of flow fields all exhibit a consistent $k^{-4}$ scaling law, and the extracted core dynamics modes (the fast central mode and the slow near-wall mode) remain highly similar across $Re$, thereby proposing a unified view of ET and EIT at low to moderate Re. Meanwhile, through Taylor-Couette flow experiments, Zhang et al. (2025) not only confirmed the continuous evolution from ET to inertia-modulated EIT but also quantitatively revealed that with increasing inertia, the velocity streaks in the flow field undergo significant tilting, and the spectral scaling laws experience a gradual transition. Choueiri et al. (2021) directly observed experimentally that as inertia increases, the key structural features of EIT undergo significant spatial migration. These findings preliminarily reveal a flow evolution paradigm characterized by elasticity dominance and inertia modulation.

Admittedly, the aforementioned studies provide important references for understanding the connection between ET and EIT; however, a systematic deficiency remains in the understanding of inertial effects. First, although linear stability analysis has found the center-mode instability across a range of $Re$, extensive high-fidelity DNS studies have not confirmed that the center mode can directly trigger EIT (Beneitez et al. 2024; Zhang et al. 2024). Moreover, while Beneitez et al. (2024b) qualitatively noted that the secondary instability resembles a center mode in ET but shifts to a wall mode in EIT—implicitly hinting at a spatial migration induced by inertia—they focused exclusively on discrete parameter points at the onset of instability, leaving the continuous modulation and spatial scaling laws of fully developed EIT completely unexplored. Second, although some studies (Rosti et al. 2024; Zhang et al. 2025) have observed a continuous transition trend from EIT to ET, their focus has been on emphasizing the equivalence of flow characteristics under low inertia, without systematically and quantitatively describing the specific modulating role of inertia. More importantly, whether it is structure identification based on modal decomposition or the qualitatively observed structural evolution from the center mode to the wall mode in experiments (Choueiri et al. 2021), there is a lack of quantitative characterization for inertial effects.

To address the above gaps, this paper systematically investigates the physical role of inertia in EIT based on DNS data of 2D viscoelastic channel flow driven by a constant bulk velocity. The present study is organized according to the following logic: First, from a macroscopic statistical perspective, we examine the overall modulating role of inertia on the flow structures of EIT. Then, we specifically discuss the amplitude and spatial modulation by inertia on statistical characteristics such as velocity, polymer extension, shear stress and energy exchange distributions in EIT. Next, based on the spatial wall-ward migration phenomenon, we derive and establish the quantitative scaling laws between characteristic physical quantities and inertial parameter, thereby defining the evolution of the critical layer. Subsequently, to explore the universality of dynamics mechanism under inertial modulation, we conduct cross-$Re$ similarity verifications on fluctuating events through probability density functions (PDFs). Finally, we reveal the universal self-sustaining mechanism by combining instantaneous flow field topology and quadrant analysis. The remainder of this paper is organized as follows: Section 2 briefly introduces the numerical methods and parameter settings; Section 3 presents detailed results and analyses, including the effects of inertia, the self-similarity and self-sustaining mechanism in EIT; Section 4 provides an in-depth discussion and draws final conclusions.
	
\section{Numerical methods and parameter settings}\label{2}
\subsection{Governing equations}

We consider the 2D plane Poiseuille flow of an incompressible viscoelastic fluid. The channel walls satisfy no-slip conditions, and periodic boundary conditions are applied in the streamwise direction. The characteristic scales are the channel half-height $h$, the bulk average velocity $u_b$, the time $h/u_b$, and the pressure $\rho u_b^2$, where $\rho$ is the fluid density. The bulk velocity is defined as $u_b = (1/2h)\int_{-h}^{h} \overline{u}(y) \, \mathrm{d}y$, with $\overline{u}(y)$ the local streamwise-averaged velocity.

The dimensionless continuity and momentum equations are

\begin{equation}
	\nabla \cdot \bm{u} = 0, \label{EQ.1}
\end{equation}

\begin{equation}
	\frac{\partial \bm{u}}{\partial t} + \bm{u} \cdot \nabla \bm{u} = -\nabla p + \frac{\beta}{Re} \nabla^2 \bm{u} + \nabla \cdot \bm{\tau}_p, \label{EQ.2}
\end{equation}

where $\bm{u} = (u, v)$ is the velocity vector, $p$ the pressure, and $\bm{\tau}_p$ the elastic stress tensor. The viscosity ratio $\beta = \eta/\mu$ relates the solvent viscosity $\eta$ to the total zero-shear viscosity $\mu$; the Reynolds number is $Re = \rho u_b h / \eta$.

The polymer solution is described by the FENE-P model. The elastic stress is expressed in terms of the conformation tensor $\bm{C}$ as

\begin{equation}
	\bm{\tau}_p = \frac{1-\beta}{Re\,Wi} \bigl[ f(r) \bm{C} - \bm{I} \bigr], \label{EQ.3}
\end{equation}

and $\bm{C}$ evolves according to

\begin{equation}
	\frac{\partial \bm{C}}{\partial t} + (\bm{u} \cdot \nabla) \bm{C} - \bm{C} \cdot (\nabla \bm{u}) - (\nabla \bm{u})^{\mathrm{T}} \cdot \bm{C} = - \frac{f(r^) \bm{C} - \bm{I}}{Wi}. \label{EQ.4}
\end{equation}

Here $\bm{C} = \langle \bm{q} \bm{q} \rangle$ with $\bm{q}$ the end-to-end vector of a polymer molecule, $Wi = \lambda u_b / h$ is the Weissenberg number ($\lambda$ the relaxation time), and $f(r) = (L^2-2)/(L^2-r^2)$ is the Peterlin function, where $r^2 = \mathrm{tr}(\bm{C})$ and $L$ is the maximum extensibility of the polymer chains.

\subsection{Numerical schemes}

The governing equations are solved using an in-house DNS code based on a staggered-grid finite-difference formulation. Spatial discretization employs a second-order central difference scheme for all terms in the momentum equation (convection, pressure, diffusion, and elastic stress). Temporal advancement uses a fractional-step method: the pressure term is treated implicitly, while the remaining terms are advanced with the second-order Adams–Bashforth scheme.

A time-splitting procedure is implemented as follows: (i) compute the conformation tensor and elastic stress from Eqs.~\ref{EQ.3} and \ref{EQ.4}; (ii) advance the velocity field partially in time to obtain a first intermediate velocity; (iii) solve the pressure Poisson equation derived from the continuity constraint to obtain a second intermediate velocity; (iv) adjust the pressure gradient to maintain a constant flow rate.

To overcome the well-known high-Weissenberg-number problem, we adopt the tensor-based interpolation method proposed in our previous work \citep{Zhang23}. This method preserves the invariants and the symmetric positive definiteness of the conformation tensor without introducing artificial diffusion, thereby enabling accurate simulations at ultra-high $Wi$. The key idea is to interpolate the eigenvalues and orientation (Euler angles) of $\bm{C}$ instead of its components.

Specifically, at a grid node $i$, the conformation tensor $\bm{C}$ is decomposed as $\bm{C} = \bm{R} \bm{\mathit{\Lambda}} \bm{R}^{\mathrm{T}}$, where the diagonal matrix of eigenvalues is $\bm{\Lambda} = \begin{bmatrix} \lambda_1 & 0 \\ 0 & \lambda_2 \end{bmatrix}$ and the rotation matrix is $\bm{R} = \begin{bmatrix} \cos\theta & -\sin\theta \\ \sin\theta & \cos\theta \end{bmatrix}$. Here $\lambda_1, \lambda_2 > 0$ are the eigenvalues, and $\theta$ is the single Euler angle (rotation angle) for 2D flows. (A 3D flow would involve a 3$\times$3 rotation matrix with three Euler angles.)

At the cell interface $i+1/2$, the eigenvalues and Euler angle are interpolated using a suitable scheme; here we employ the weighted essentially non-oscillatory (WENO) scheme \citep{Shu98} for the convective term in Eq.~\ref{EQ.4}. Once the interpolated $\bm{\mathit{\Lambda}}_{i+1/2}$ and $\bm{R}_{i+1/2}$ are reconstructed, the conformation tensor at the interface is obtained as $\bm{C}_{i+1/2} = \bm{R}_{i+1/2} \bm{\mathit{\Lambda}}_{i+1/2} \bm{R}_{i+1/2}^{\mathrm{T}}$. Further details can be found in \citet{Zhang21a, Zhang23, Zhang24}.

\subsection{Numerical conditions}

In our previous work \citep{Zhang24}, it was reported that the computational domain length used in viscoelastic channel flow simulation significantly affects the flow state. Particularly for EIT research, we avoided this problem by conducting multiple simulations using channel with different lengths. Nevertheless, we still presented numerous cases, with parameter settings, corresponding channel lengths and different flow states, as shown in Table~\ref{tab1}. In this study, the channel flow direction length $L_x$ ranges from 10 to 40$h$ and the wall-normal length $L_y$ is 2$h$. The grid is uniform in the streamwise direction and clustered near the walls in the wall-normal direction. The wall-normal grid distribution is as follows:

\begin{equation}
	y_j = \frac{1}{a} \tanh\left( \frac{1}{2} \delta_j \ln\frac{1+a}{1-a} \right), \quad \delta_j = -1 + \frac{2j}{N_y}, \label{EQ.5}
\end{equation}
with $a = 0.95$ controlling the clustering. Following grid-independence tests \citep{Zhang24, Cheng25}, the resolution is set to $N_x \times N_y = 512n \times 304$ ($n=L_x/10h$), and the time step setting is shown in Table~\ref{tab1}. All simulations are performed at fixed $\beta = 0.9$, and $L = 100$, while $Re$ varies from 100 to 6000 and $Wi$ varies from 2 to 200. Table~\ref{tab1} summarizes the numerical parameters and selected results. The friction Reynolds number is $Re_\tau = u_\tau h/\nu$ and the friction Weissenberg number is $Wi_\tau = \lambda u_\tau/h$, where $u_\tau = \sqrt{\tau_w/\rho}$ is the friction velocity. The friction coefficient $C_f$ is averaged over long simulation times in the longer channel.

\begin{table}
	\centering
	\caption{Partial DNS results and corresponding parameter settings. $Re_\tau = u_\tau h/\nu$, $Wi_\tau = \lambda u_\tau/h$. The average friction coefficient $C_f$ is the spatial and temporal average of long-time simulation results in the longer channel.}
	
\label{tab1}
\begin{tabular}{cccccccc}
	\toprule
	$Wi$ & $Re$ & $Wi_\tau$ & $Re_\tau$ &  $L_x$  & $\Delta t$ ($10^4 \times h/u_b$) & $C_f$ & Flow state \\
	\midrule
	100 & 500 & 335.75 & 40.97 & 20 / 40 & 1.25 & 0.01343 & SAR / EIT \\
	100 & 700 & 368.20 & 50.77 & 20 / 40 & 1.75 & 0.01052 & EIT \\
	100 & 1000 & 417.00 & 64.58 & 20 / 40 & 2.50 & 0.00834 & EIT \\
	100 & 2000 & 507.89 & 100.79 & 20 & 5.00 & 0.00508 & EIT \\
	100 & 4000 & 622.44 & 157.79 & 10 / 20 & 10.00 & 0.00311 & EIT \\
	100 & 6000 & 689.66 & 203.42 & 10 / 20 & 10.00 & 0.00230 & EIT \\
	40 & 700 & 129.77 & 47.66 & 20 / 40 & 1.75 & 0.00927 & CAR / EIT \\
	40 & 1000 & 145.94 & 60.40 & 20 / 40 & 2.50 & 0.00730 & EIT / CAR \\
	40 & 2000 & 176.80 & 94.02 & 20 & 5.00 & 0.00442 & EIT \\
	40 & 4000 & 225.60 & 150.20 & 10 / 20 & 10.00 & 0.00282 & EIT \\
	40 & 6000 & 256.80 & 196.27 & 10 / 20 & 10.00 & 0.00214 & EIT \\
	20 & 700 & 61.13 & 46.25 & 20 / 40 & 1.75 & 0.00873 & EIT \\
	20 & 1000 & 65.09 & 57.05 & 10 / 20 & 2.50 & 0.00651 & EIT \\
	20 & 2000 & 79.20 & 88.99 & 20 & 5.00 & 0.00396 & EIT \\
	20 & 4000 & 101.73 & 142.64 & 10 / 20 & 10.00 & 0.00254 & EIT \\
	20 & 6000 & 117.91 & 188.07 & 10 / 20 & 10.00 & 0.00197 & EIT \\
	3.3 & 333 & 9.99 & 31.59 & 20 & 1.00 & 0.01800 & L \\
	16.7 & 333 & 49.15 & 31.34 & 20 & 1.00 & 0.01771 & L \\
	20 & 333 & 58.74 & 31.27 & 20 & 1.00 & 0.01764 & L \\
	26.7 & 333 & 77.32 & 31.18 & 20 & 1.00 & 0.01756 & SAR \\
	30 & 333 & 86.36 & 31.12 & 20 & 1.00 & 0.01754 & SAR \\
	5.3 & 533 & 15.95 & 39.92 & 20 & 1.50 & 0.01122 & L \\
	10.7 & 533 & 31.70 & 39.80 & 20 & 1.50 & 0.01115 & EIT \\
	16 & 533 & 47.33 & 39.71 & 20 & 1.50 & 0.01110 & EIT \\
	21.3 & 533 & 62.60 & 39.55 & 20 & 1.50 & 0.01101 & L \\
	26.7 & 533 & 77.82 & 39.44 & 20 & 1.50 & 0.01095 & SAR \\
	32.0 & 533 & 93.13 & 39.38 & 20 & 1.50 & 0.01092 & CAR \\
	42.7 & 533 & 139.29 & 41.81 & 20 & 1.50 & 0.01225 & CAR \\
	6.7 & 667 & 19.92 & 44.64 & 20 & 1.70 & 0.00896 & L \\
	20 & 667 & 60.63 & 44.97 & 20 & 1.70 & 0.00909 & EIT \\
	26.7 & 667 & 82.89 & 45.53 & 20 & 1.70 & 0.00932 & EIT \\
	33.3 & 667 & 106.94 & 46.26 & 20 & 1.70 & 0.00962 & CAR \\
	40 & 667 & 132.20 & 46.95 & 20 & 1.70 & 0.00991 & CAR \\
	53.3 & 667 & 181.42 & 47.63 & 20 & 1.70 & 0.01020 & CAR \\
	60 & 667 & 207.70 & 48.05 & 20 & 1.70 & 0.01038 & CAR \\
	10 & 1000 & 29.80 & 54.59 & 10 / 20 & 2.50 & 0.00596 & EIT \\
	60 & 1000 & 228.00 & 61.64 & 20 & 2.50 & 0.00760 & EIT \\
	9.5 & 1333 & 28.49 & 63.23 & 20 & 3.00 & 0.00450 & L \\
	13.3 & 1333 & 42.74 & 65.37 & 20 & 3.00 & 0.00481 & EIT \\
	40 & 1333 & 156.80 & 72.30 & 20 & 3.00 & 0.00588 & EIT \\
	66.7 & 1333 & 281.26 & 74.99 & 20 & 3.00 & 0.00633 & EIT \\
	120 & 1333 & 562.26 & 79.03 & 20 & 3.00 & 0.00703 & EIT \\
	8.9 & 2000 & 26.52 & 77.20 & 20 & 5.00 & 0.00298 & L \\
	10 & 2000 & 32.70 & 80.87 & 20 & 5.00 & 0.00327 & EIT \\
	60 & 2000 & 285.60 & 97.57 & 20 & 5.00 & 0.00476 & EIT \\
	120 & 2000 & 627.60 & 102.27 & 20 & 5.00 & 0.00523 & EIT \\
	150 & 2000 & 804.00 & 103.54 & 20 & 2.50 & 0.00536 & EIT \\
	180.0 & 2000 & 986.40 & 104.69 & 20 & 2.50 & 0.00548 & EIT \\		
	200 & 2000 & 1116.00 & 105.64 & 20 & 2.50 & 0.00558 & EIT \\
	\bottomrule
\end{tabular}
\end{table}

	\section{Results and analysis}\label{3}
	\subsection{Flow State Transition and Topological Structure Evolution}
		
\begin{figure}
	\centering
	\includegraphics[width=1\textwidth]{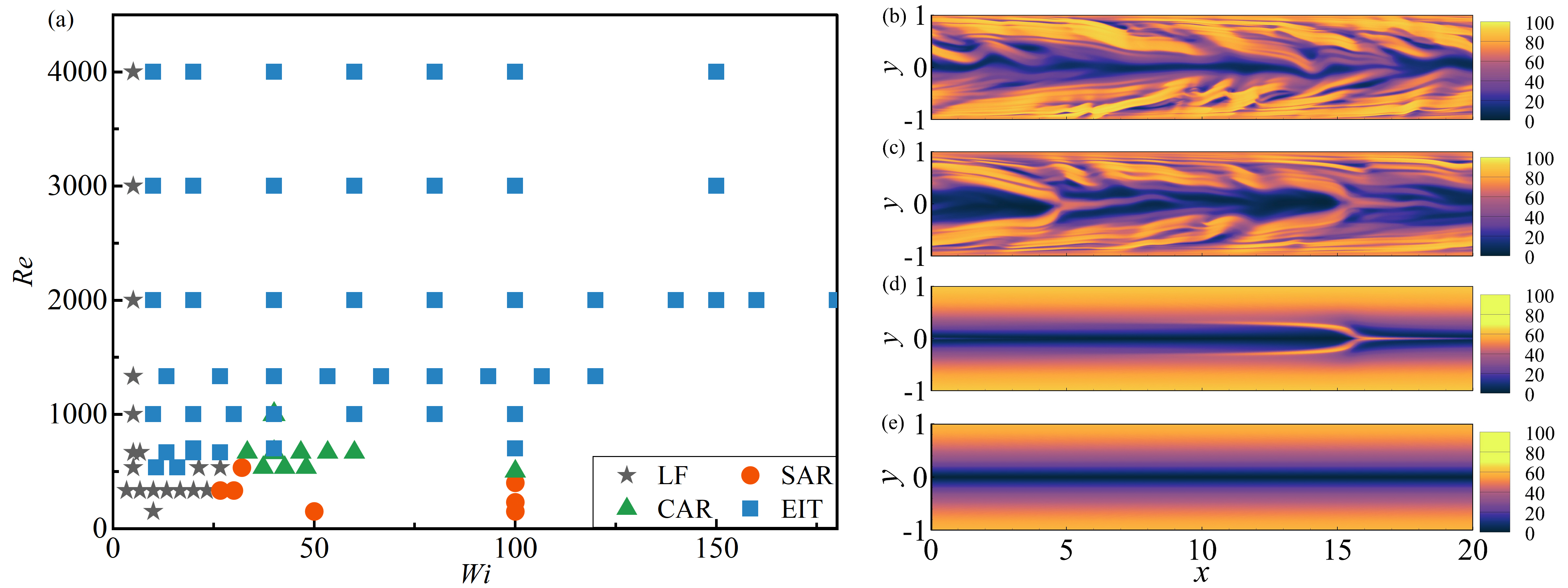}
	\caption{\label{fig1} Phase diagram (a) and schematic diagrams (b-e) of different flow states.}
\end{figure}

Figure \ref{fig1} systematically presents the regime map for 2D viscoelastic channel flows across a wide parameter space $(Wi, Re)$. Based on extensive direct numerical simulations (DNS), the phase diagram (figure \ref{fig1}a) identifies four self-sustaining states: laminar flow (LF), steady arrowhead regime (SAR), chaotic arrowhead regime (CAR), and EIT. The corresponding instantaneous polymer stretching fields (figure \ref{fig1}b--e) illustrate the fundamental topological differences between these states. EIT is characterized by spatiotemporally chaotic, large-scale stretching sheets distributed throughout the channel. While CAR is also chaotic and exhibits statistical properties nearly identical to EIT away from the centerline (Beneitez et al., 2023), it persistently retains a visual signature of a weak, distorted central arrowhead structure. Conversely, SAR represents a nonlinearly traveling wave, manifesting as a symmetric arrowhead structure propagating at a constant speed. Notably, the number of these structures depends on the parameters and domain size; for instance, at $Re=400$ and $Wi=100$, three arrowheads can coexist in a $L_x=40$ channel (not shown).
The transitions between these flow states reflect the competition between two intrinsic dynamical mechanisms: wall mode and center mode. The former has been demonstrated as the origin of the chaotic dynamics in EIT (Zhang et al., 2024; Beneitez et al., 2024), manifesting as elongated stretching structures near the wall. The latter, driven by strong elasticity, reaches a nonlinear saturation in the form of arrowhead structures (Page et al., 2020; Dubief et al., 2022). The dominance of either mechanism is modulated by a combination of inertia, elasticity, the computational domain size etc.

First, the channel streamwise length significantly impacts the realized flow state. We previously explored the influence of domain size in the minimal flow unit (MFU) context (Zhang et al., 2024). At $Re=2000$ and $Wi=100$, increasing the channel length from $L_x=5$ to $10$ or $20$ triggers a transition from SAR to EIT. Similarly, for $Re=700, Wi=40$, doubling $L_x$ from $20$ to $40$ leads to a transition from CAR to EIT, suggesting that larger domain sizes may weaken the effect of center mode. Paradoxically, at $Re=1000, Wi=40$, increasing $L_x$ from $10$ to $20$ results in a regression from EIT to CAR. This contradictory behavior highlights the unpredictable sensitivity of modal competition to boundary conditions within certain parameter regimes. The underlying mechanism by which domain size alters the flow states remains an open question, requiring further numerical investigation.
Second, the evolution pathways of these states under varying parameters clearly reflect the shifting balance between the above mechanisms. Under low-elasticity conditions ($Wi=20$), decreasing $Re$ leads to a direct decay from EIT to laminar flow without intermediate states. However, at higher elasticity ($Wi=100$), a complete degeneration process is observed: EIT $\rightarrow$ CAR $\rightarrow$ SAR. This implies that EIT clearly originates from the wall mode instabilities at low elasticity, and insufficient inertia will lead to complete attenuation; when elasticity are sufficiently strong, the center modes can become unstable, and inadequate inertia will result in EIT being replaced by the arrowhead structures of center mode.
Of particular interest is the observation that at high elasticity, the SAR state arising from inertia lack is topologically almost identical to the "Narwhal" structures reported in recent studies of ET (Morozov 2022; Morozov et al. 2025). As shown in figure \ref{fig1}(a), at $Wi=100$, the arrowhead state persists even as $Re$ is reduced to $400$ or even $100$, where inertia is profoundly weak. Under this parameter ( $Re = 100 - 400$) of current channel flow, no turbulent states have been obtained. Even the reported ET in channel flow occurs under Re<<1 (Morozov et al. 2025; Lin et al. 2025). However, this precisely demonstrates the universal presence of such regular structures in broad-parameter spaces (Re range spanning 2 to 4 orders of magnitude), and the structures are highly likely to serve as a bridge within the transition regions between ET and EIT. Within this framework, Inertia acts as a 'sliding resistor' on the flow state, with its strength determining the switching between EIT and ET. Thus, the formation and self-sustenance of EIT rely on a requisite level of inertia, below which the flow exhibit laminarization, SAR state, or other complex transitional forms. This highlights the importance of investigating the role of inertia in EIT.

\begin{figure}
	\centering
	\includegraphics[width=0.6\textwidth]{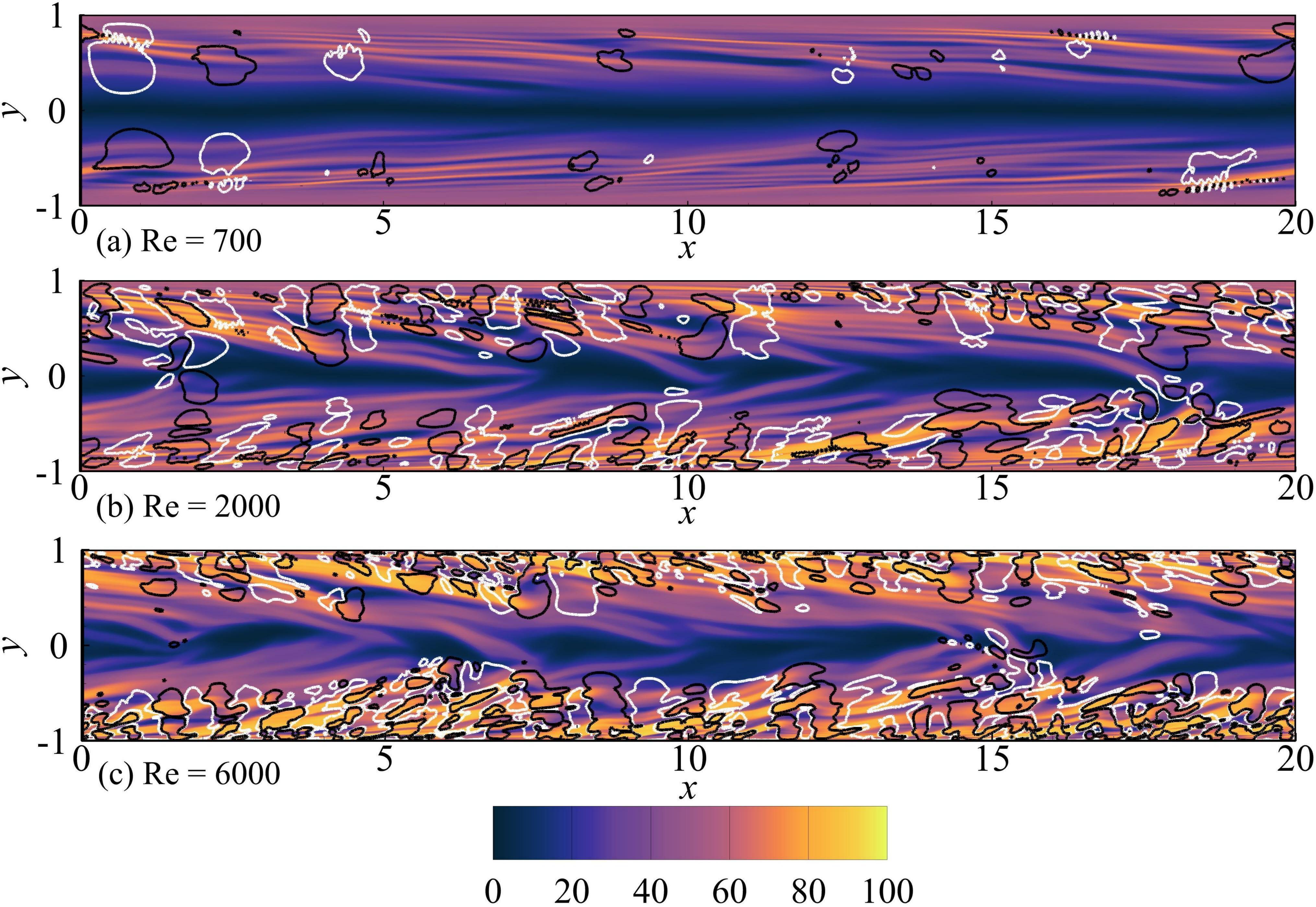}
	\caption{\label{fig2} Variation of vortex structure with Re under Wi = 20 (Q = ± 0.02).}
\end{figure}

\begin{figure}
	\centering
	\includegraphics[width=1\textwidth]{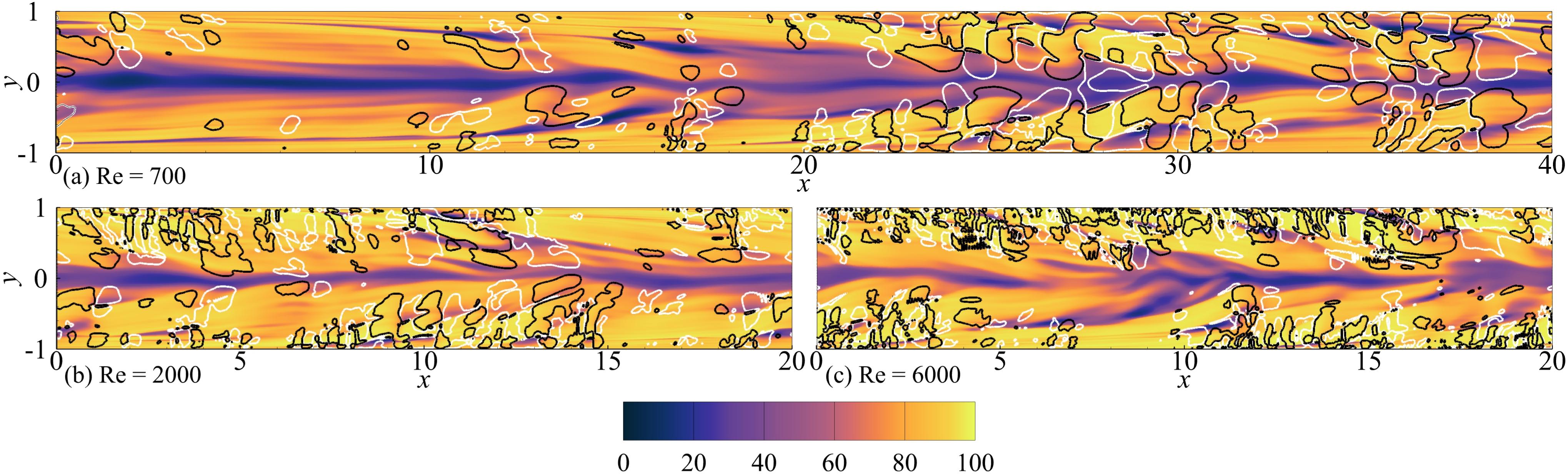}
	\caption{\label{fig3} Variation of vortex structure with Re under Wi = 100 (Q = ± 0.02).}
\end{figure}

To intuitively reveal how inertial effects reshape the topological structures of the flow, figures \ref{fig2} and \ref{fig3} display the instantaneous polymer stretching fields at various Reynolds numbers for $Wi = 20$ and $100$, respectively. The $Q$-criterion contours ($Q = \pm 0.02$) are superimposed to identify the vortical features. Regarding the intensity distribution of the stretching field, an increase in the Reynolds number significantly promotes the magnitude of polymer stretching, indicating that inertia and elasticity share a similar enhancing trend in intensifying EIT. For instance, in the flow field at $Wi = 20$ and $Re = 700$ shown in figure \ref{fig2}(a), the exhibited quasi-instability structures are highly localized and weak in spatial distribution, failing to spread throughout the entire domain (this flow state is likely a transitional or under-developed regime). With an evident increase in $Re$, the turbulence intensity intensifies, driving the chaotic stretching structures to fully diffuse throughout the entire flow field.

However, a more striking variation is manifested in the topological evolution of vortical structures. Under both low and high $Wi$ conditions, the flow fields at low $Re$ ($Re = 700$) are dominated by large-scale, sparsely distributed vortices that can extend into the core region far from the wall. This is evidently attributed to the suppression of small-scale structures by strong viscous diffusion at low $Re$. As Re increases, the nonlinear inertia effects are significantly enhanced, causing large-scale vortices to undergo remarkable fragmentation, disintegrating into fine and densely packed clusters of micro-vortices. Concurrently, these small-scale vortical structures exhibit a strong near-wall affinity in physical space, migrating from the channel center to the near wall. The structural fragmentation and wall-trending migration is fundamentally originates from the thinning of shear boundary layer at high $Re$.
Therefore, in 2D EIT, the inertia does not merely amplify the turbulence intensity; rather, it appears to comprehensively modulate EIT by altering the near-wall shear environment, encompassing promoting the enhancement of dynamics amplitude and the wall-ward migration of core structures. To further elucidate the specific role of inertia in 2D EIT, we will proceed with detailed quantitative analyses in the following sections.

	\subsection{Statistical Characteristics Evolution}
     
\begin{figure}
	\centering
	\includegraphics[width=0.6\textwidth]{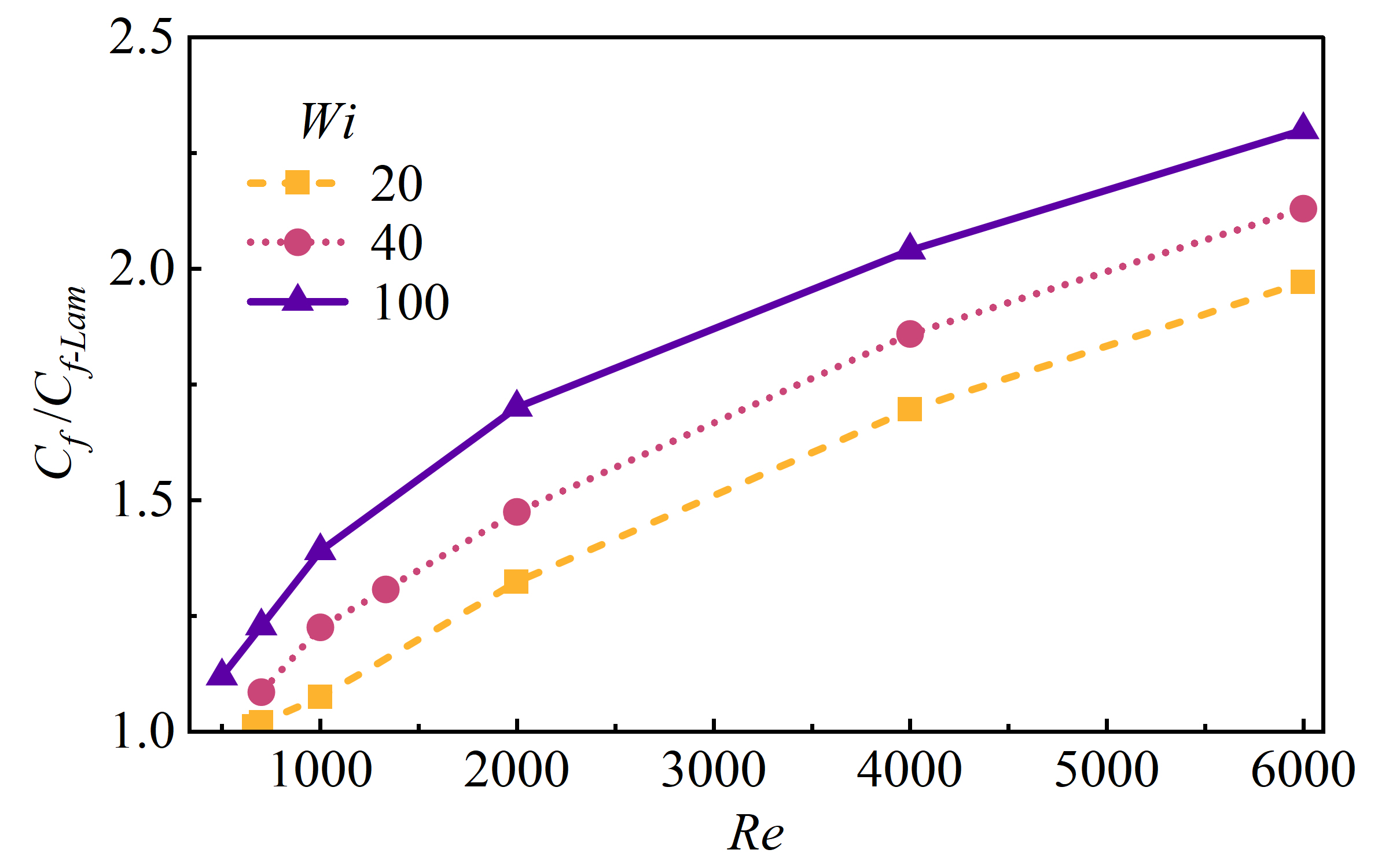}
	\caption{\label{fig4} Growth rate of friction coefficient with Re for different Wi cases.}
\end{figure}

\begin{figure}
	\centering
	\includegraphics[width=0.7\textwidth]{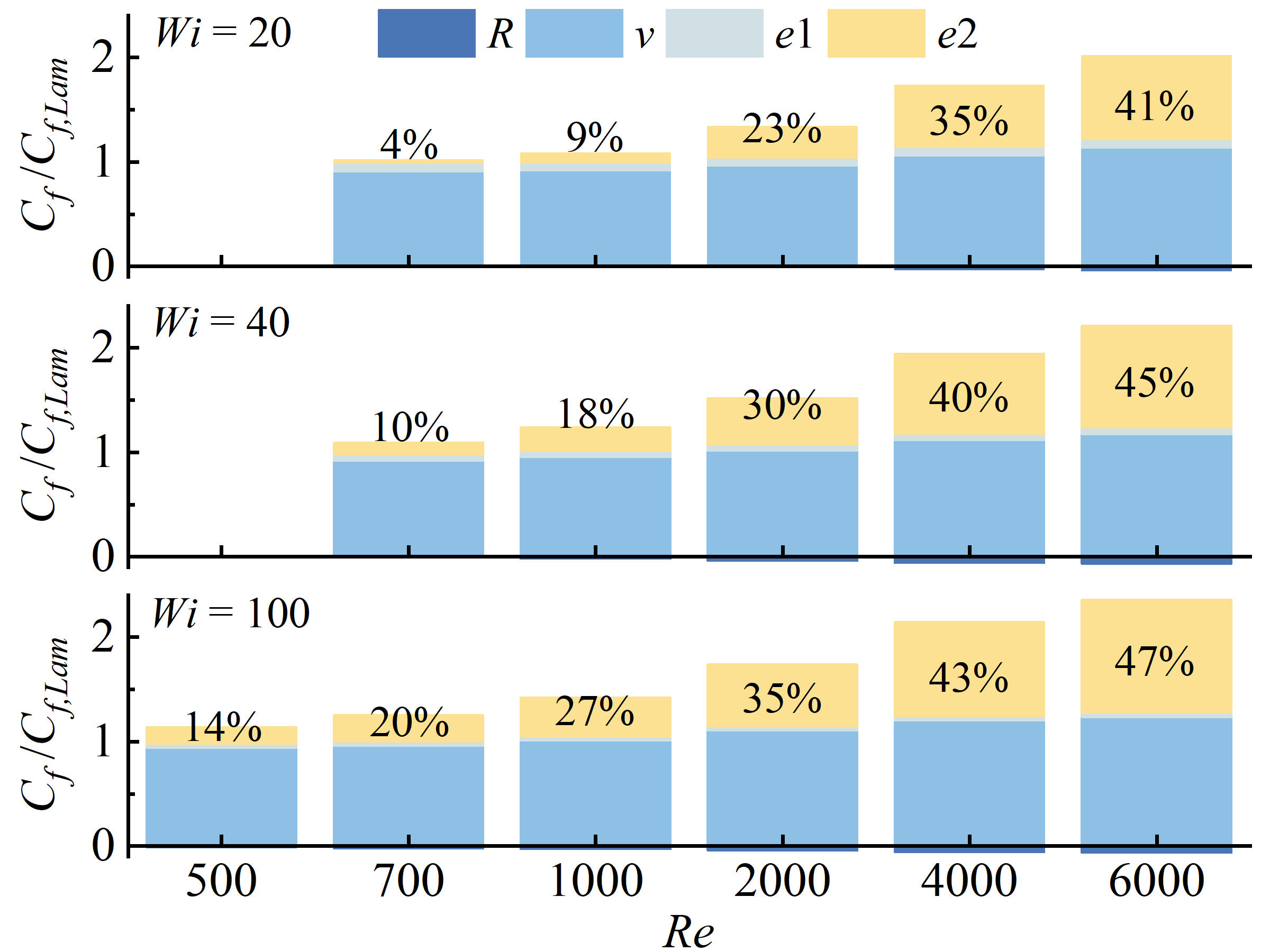}
	\caption{\label{fig5} Contributions to the flow drag coefficient under different Re and Wi based on the R-D identity.}
\end{figure}

The average friction coefficient ($C_f$) acts as a fundamental macroscopic indicator of flow states. Figure \ref{fig4} presents the ratio of $C_f$ to the equivalent Newtonian laminar $C_f$ at the same $Re$, characterizing the drag enhancement rate. As $Re$ increases, this ratio exhibits an overall upward trend, quantificationally corroborating the previous observation: the inertia significantly amplifies the turbulent intensity of EIT. To trace the dynamical origin of drag enhancement, we employ the Renard--Deck (RD) identity (Zhang et al., 2020) to decompose the average friction coefficient into spatial integrals of distinct stress components. As shown in figure \ref{fig5}, the total drag is exactly divided into four contributions: viscous shear stress, linear elastic shear stress, nonlinear elastic shear stress, and Reynolds shear stress. The viscous shear stress persistently dominates the total drag, while the absolute contributions from the linear elastic and Reynolds shear stresses remain marginal. Notably, the Reynolds shear stress provides a slight negative contribution, which stems from the anomalous negative Reynolds stress characteristic inherent to 2D EIT, a phenomenon we detailed in our prior work (Cheng et al., 2025). The most critical finding here is the substantial growth in the proportion of the nonlinear elastic shear stress contribution driven by increasing inertia. It is known that enhanced elasticity (increasing $Wi$) directly elevates this nonlinear term (Wang et al., 2023); however, our data indicate that although nonlinear elastic stress exhibits elastic characteristics, it originates from polymer stretching. The significant promotion of inertia to polymer stretching as demonstrated in Figure 2 and 3, is naturally evident on the nonlinear $\tau_e$.

\begin{figure}
	\centering
	\includegraphics[width=1\textwidth]{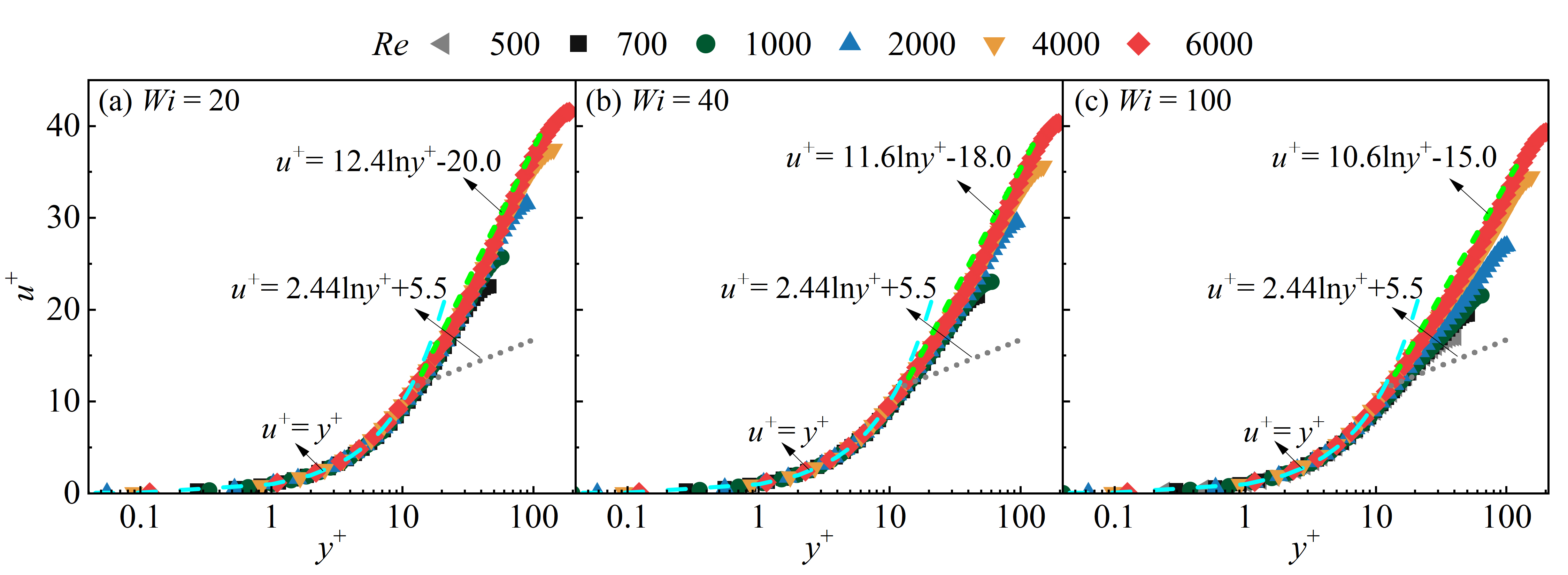}
	\caption{\label{fig6} Mean streamwise velocity profiles based on the inner scale.}
\end{figure}

The role of inertia is also directly manifested in the velocity field. Figure \ref{fig6} displays the inner-scaled mean streamwise velocity profiles, $u^+$, for $Wi = 20$, $40$, and $100$. At a fixed $Wi$, the velocity profiles across the channel region continuously shift upward and tend toward asymptotic saturation as $Re$ increases. First, we note that this upward shift is opposite to the trend observed when increasing $Wi$ at a fixed $Re$ (Cheng et al., 2025). This discrepancy primarily originates from the disparate evolutionary behaviors of the friction coefficient. It must be emphasized that this upward shift essentially reflects the fundamental kinematic scaling whereby the absolute friction coefficient $C_f$ naturally decays with increasing $Re$. However, despite the opposing trends in the mean velocity profiles, the enhancement of either inertia or elasticity leads to an increased proportional contribution from the nonlinear elastic shear stress (as shown in figures \ref{fig4} and \ref{fig5}). Consequently, this upward shift of the velocity profile does not imply a reversion towards laminar flow; rather, it indicates that the flow is evolving toward a fully developed 2D EIT. Secondly, the asymptotic saturation of the velocity profiles implies the emergence of a logarithmic region. However, we find that the slope of this logarithmic region is still significantly modulated by $Wi$, hence a specific quantitative description is not provided here. Concurrently, this saturation behavior strongly hints at the existence of a certain statistical self-similarity in the fully developed stage, a subject that will be explored in depth in the subsequent sections.

\begin{figure}
	\centering
	\includegraphics[width=1\textwidth]{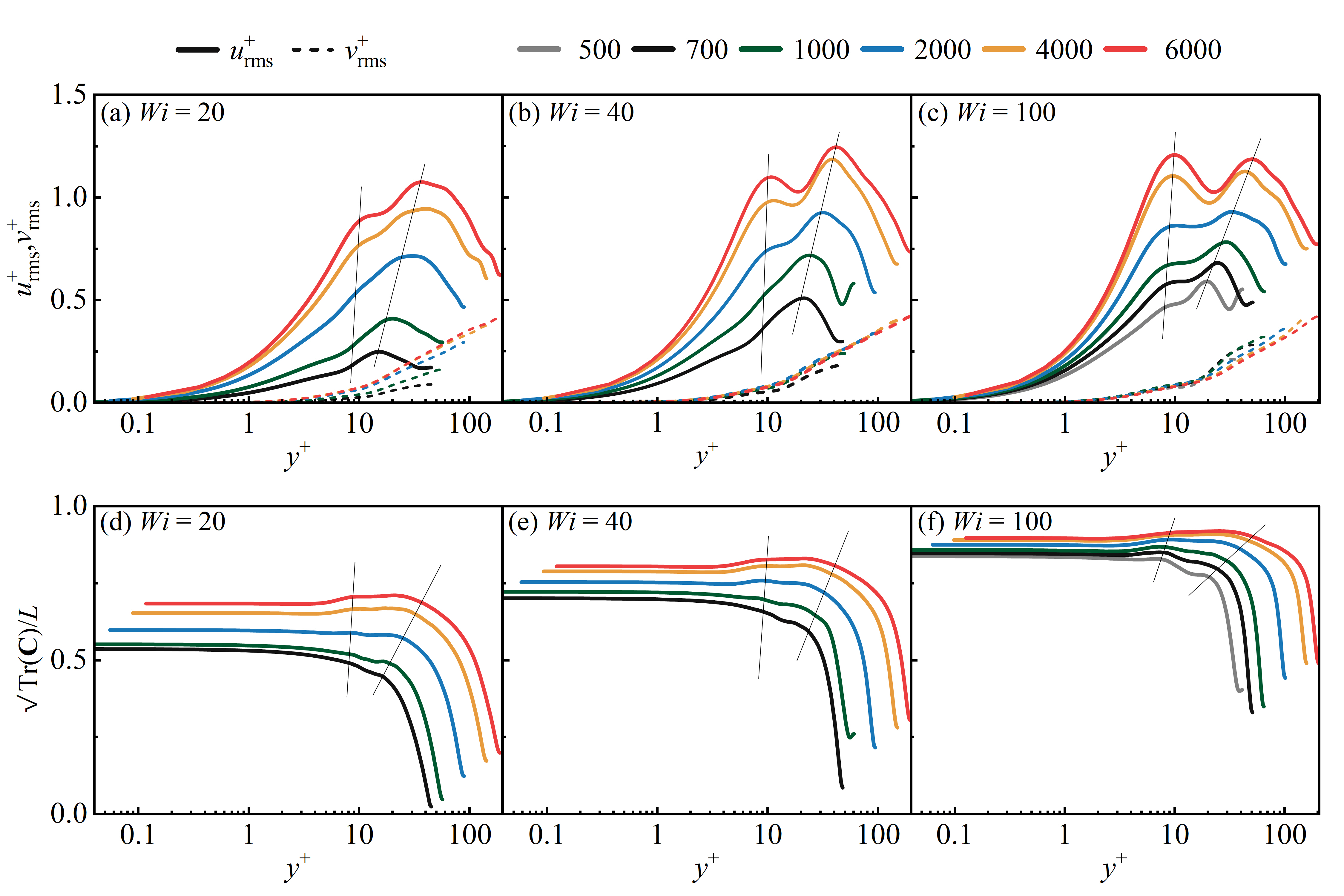}
	\caption{\label{fig7} Velocity fluctuations root mean square and average extension profiles based on internal scale normalization.}
\end{figure}

Furthermore, figure \ref{fig7} illustrates the inner-scaled wall-normal ($y^+$) distributions of velocity fluctuations root-mean-square (RMS) and mean polymer extension. Figures \ref{fig6}(a-c) correspond to the streamwise ($u_{\mathrm{rms}}$, solid lines) and wall-normal ($v_{\mathrm{rms}}$, dashed lines) velocity RMS at $Wi$ = 20, 40, and 100, respectively. The $u_{\mathrm{rms}}$ profiles exhibit a monotonic enhancement with increasing $Re$. Conversely, the evolution of $v_{\mathrm{rms}}$ is highly sensitive to $Wi$, transitioning from a monotonic increase to an increase-then-decrease pattern, and eventually to a monotonic decrease at high $Wi$. This divergence suggests a dynamic competition between inertia and elasticity in wall-normal momentum transfer: at lower elasticity, enhanced inertia assists the elasticity to dominate the flow state; whereas when elasticity is sufficiently dominant, the enhanced inertia exerts a suppressive effect. Despite the complex parametric dependence of $v_{\mathrm{rms}}$, its magnitude remains consistently and significantly lower than $u_{\mathrm{rms}}$, highlighting that the core fluctuation energy of EIT is concentrated in the streamwise direction, exhibiting strong anisotropy. It should be noted that while some cases (Wi=40 and Re=1000, Wi=100 and Re=500) manifest a CAR state, causing extra fluctuation peaks at the channel center, their statistical properties outside the central region are almost indistinguishable from standard EIT (Dubief et al. 2022), thus unaffecting our core analysis of non-central regions.

Particularly noteworthy is the emergence of dual-peak distribution in the velocity RMS, prominently visible in $u_{\mathrm{rms}}$. In figures \ref{fig7}(a-c), the near-wall region of $u_{\mathrm{rms}}$ initially shows a slight inflection point at low parameters. As $Re$ and $Wi$ increase, this inflection gradually protrudes and evolves into a distinct secondary peak alongside the overall amplitude amplification. We propose two physical hypotheses for the origin of this bimodal phenomenon. First, analogizing the central peak induced by the unstable center mode in the CAR state, we hypothesize that the dual peaks of $u_{\mathrm{rms}}$ in 2D EIT are triggered by two different non-center modes. Such multi-peak structures are inexistent in traditional Newtonian turbulence and 3D MDR, underscoring the unique dynamics of 2D EIT. Second, a topological evidence is provided by the mean polymer extension field. Figures \ref{fig7}(d-f) show the normalized mean polymer extension $\sqrt{Tr(C)/L^2}$. Similar to $u_{\mathrm{rms}}$, the mean extension is positively promoted by inertia, consistent with the instantaneous contours in figure \ref{fig2} and \ref{fig3}. Crucially, at wall-normal locations approximately corresponding to the dual peaks of $u_{\mathrm{rms}}$, the mean extension profiles exhibit two distinct, non-smooth kinks. By correlating these with the instantaneous topology (figure \ref{fig2}), we posit that these two kinks spatially correspond to the leading and trailing edges of the highly frequent, sheet-like stretched polymer structures. The two intense local gradient disparities in stretching frequency inevitably feedback into the momentum equation, exciting local velocity fluctuations that manifest as the bimodal $u_{\mathrm{rms}}$ distribution.

\begin{figure}
	\centering
	\includegraphics[width=1\textwidth]{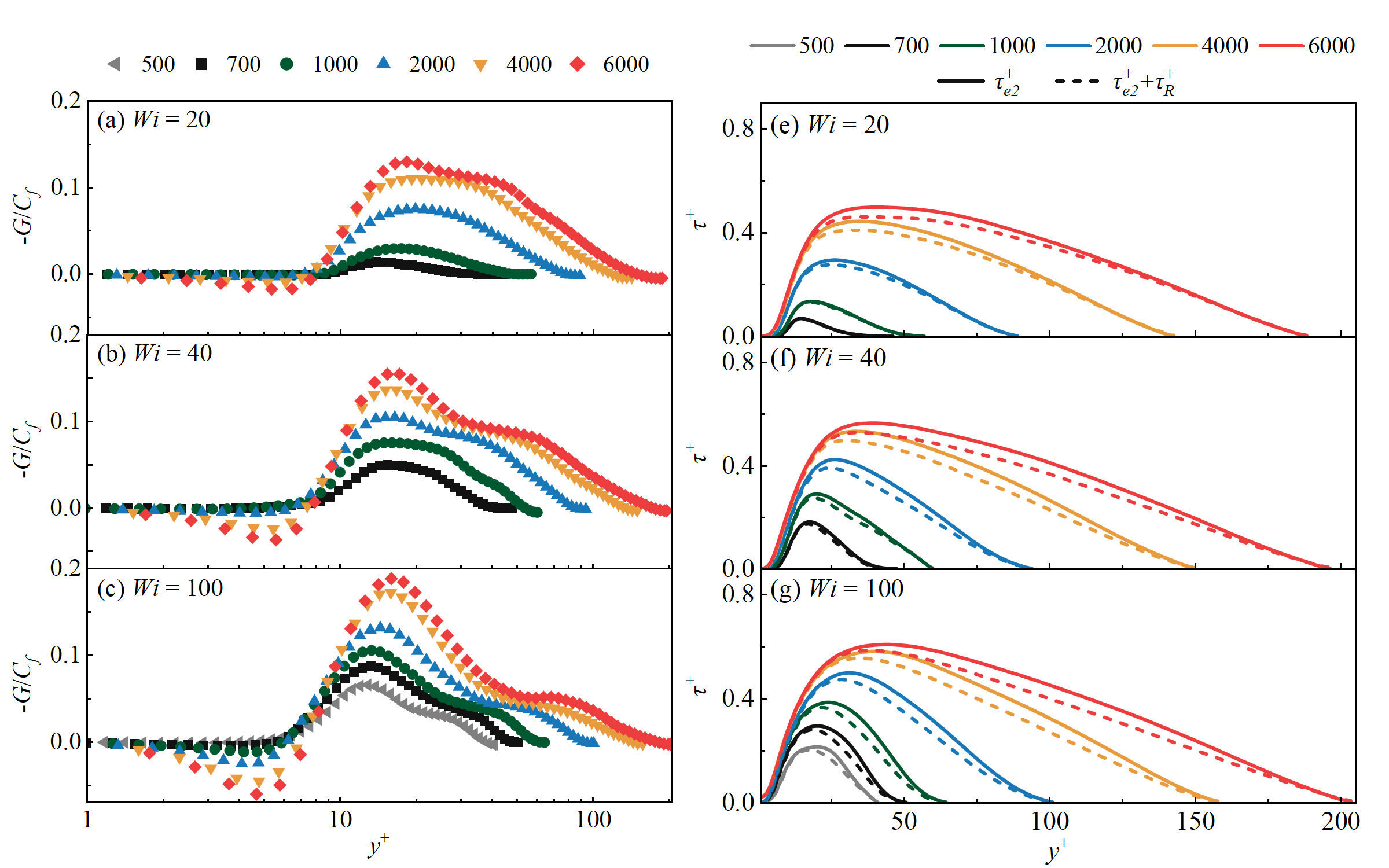}
	\caption{\label{fig8} Evolution of (a) the ratio profiles between energy exchange term and friction coefficient and (b) nonlinear elastic shear stress (nonlinear elastic shear stress + Reynolds stress) profiles with Re across different Wi.}
\end{figure}

In addition to velocity fluctuations and mean ploymer extension, the amplifying effect of inertia on EIT intensity is pervasively reflected across various statistics. Figure \ref{fig8} display the $Re$-evolution of the energy exchange rate ($-G/C_f$) and the inner-scaled nonlinear elastic shear stress profiles, respectively. Both quantities exhibit monotonic amplitude surges with increasing $Re$, perfectly corresponding to the intensification of extension structures with elastic nature observed in physical space. It must be emphasized that the underlying mechanism for turbulent kinetic energy (TKE) enhancement in 2D EIT differs fundamentally from Newtonian turbulence. In Newtonian flows, the Reynolds stress ($-u'v'$) directly dominates TKE production. In EIT, however, the enhanced inertia intensifies the near-wall shear, thereby elevating the friction Wi ($Wi_\tau$). This high-$Wi_\tau$ environment forces the polymer chains into extreme stretching, storing massive elastic energy, which is subsequently and efficiently converted into TKE fluctuations during relaxation. Thus, even though the energy exchange rate and elastic shear stress are fundamentally elastic properties, they are strongly and indirectly driven by inertia within the self-sustaining cycle of EIT.

\begin{figure}
	\centering
	\includegraphics[width=1\textwidth]{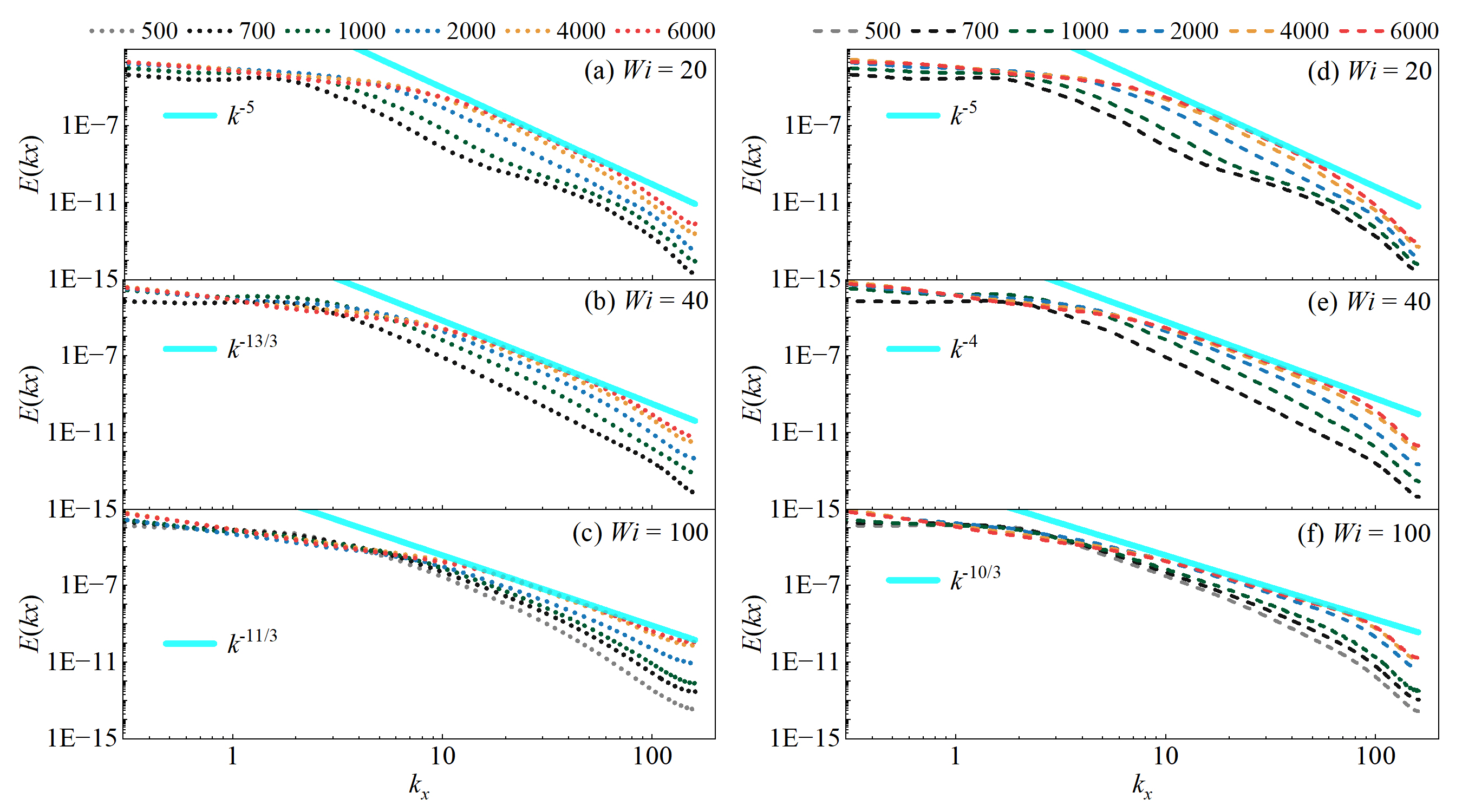}
	\caption{\label{fig9} Evolution of turbulent kinetic energy spectrum with Re at (a) $-G/C_f$ and (b) nonlinear elastic shear stress peaks across different Wi.}
\end{figure}

Given the sharp, prominent single peaks in figure \ref{fig8}, which precisely pinpoint the locations of the most intense energy conversion and elastic stress, the spectral characteristics of TKE attenuation at these specific locations are highly representative of the overall dynamics. As shown in figure \ref{fig9}, the TKE spectra for all cases exhibit a power-law decay far steeper than the Newtonian inertial subrange ($-5/3$), falling between $-5$ and $-10/3$. This aligns well with the spectral signature characteristic of viscoelastic turbulence (Fouxon et al., 2003). More importantly, for a fixed $Wi$, as $Re$ increases, the spectral curves gradually flatten in the high-wavenumber region, with the decay exponent asymptotically converging to a specific value. This statistical trend in frequency space serves as a counterpart to the physical-space vortex evolution (figures \ref{fig2} and \ref{fig3}): driven by high inertia, the quasi-regular vortex structures break down into fragmented and densely packed small-scale vortex clusters, directly resulting in a relative increase in the energy fraction at high frequencies.

In summary, both the flow structures and the spatio-temporally averaged statistical characteristics consistently demonstrate that inertia plays an indispensable modulating role in 2D EIT. This modulation is primarily manifested as a significant amplification of dynamical magnitudes and a wall-ward migration of their spatial locations. Therefore, it is necessary to quantitatively describe this modulation effect. It should be noted that the surge in the magnitudes of these statistics is subjected to a complex joint modulation by both inertia and elasticity. Similar to how the slope of the log-law in the mean velocity profile is intricately influenced by $Wi$, making it difficult to determine an exact analytical value, extracting a scaling law for amplitude variations solely with respect to inertia is challenging to implement. In contrast, the wall-ward migration is predominantly regulated by inertia; thus, conducting a quantitative scaling analysis on this spatial shift is both feasible and necessary, which constitutes the focus of Section 3.3. Meanwhile, does this strong dependence of profile amplitudes and peak locations on inertia imply that the underlying dynamical mechanisms of EIT are also altered? In Section 3.4, we will delve into the probability space by analyzing the probability density functions (PDFs) of the normalized fluctuations, aiming to explore the hidden statistical self-similarity and self-sustaining mechanism.

	\subsection{Scale Law Relationship Between Critical Layer and Inertial Effects}

Beyond the aforementioned observations, a core phenomenon of interest is the wall-ward migration of core structures in EIT driven by inertia. Statistically, this manifests as the peak locations of physical quantities shifting to larger $y^+$ as $Re$ increases (which corresponds to a migration toward the wall in outer scale). To reveal the role of inertia, we systematically track the evolution of the peak locations ($y^+$) with respect to $Re_\tau$. Since the velocity fluctuations RMS profiles exhibit dual peaks or inflection points that make maximum point extraction ambiguous, we restrict our systematic investigation to the distributions of $-G/C_f$ and $\tau_e$. Meanwhile, apart from the TKE spectrum attenuation indices (-14/3, -4, -19/6) and the quantitative relationships between MFU size, friction coefficient, and Wi derived from our previous work (Zhang et al. 2024), existing scaling relationships are predominantly confined to idealized linear stability analyses (Garg et al. 2018; Khalid et al. 2021b; Wan et al. 2023). Consequently, quantitative descriptions based on DNS or experimental data are severely lacking in EIT-related studies. Fortunately, the following will address this gap by establishing scaling law relationships to characterize the influence of inertial effects.

\begin{figure}
	\centering
	\includegraphics[width=0.6\textwidth]{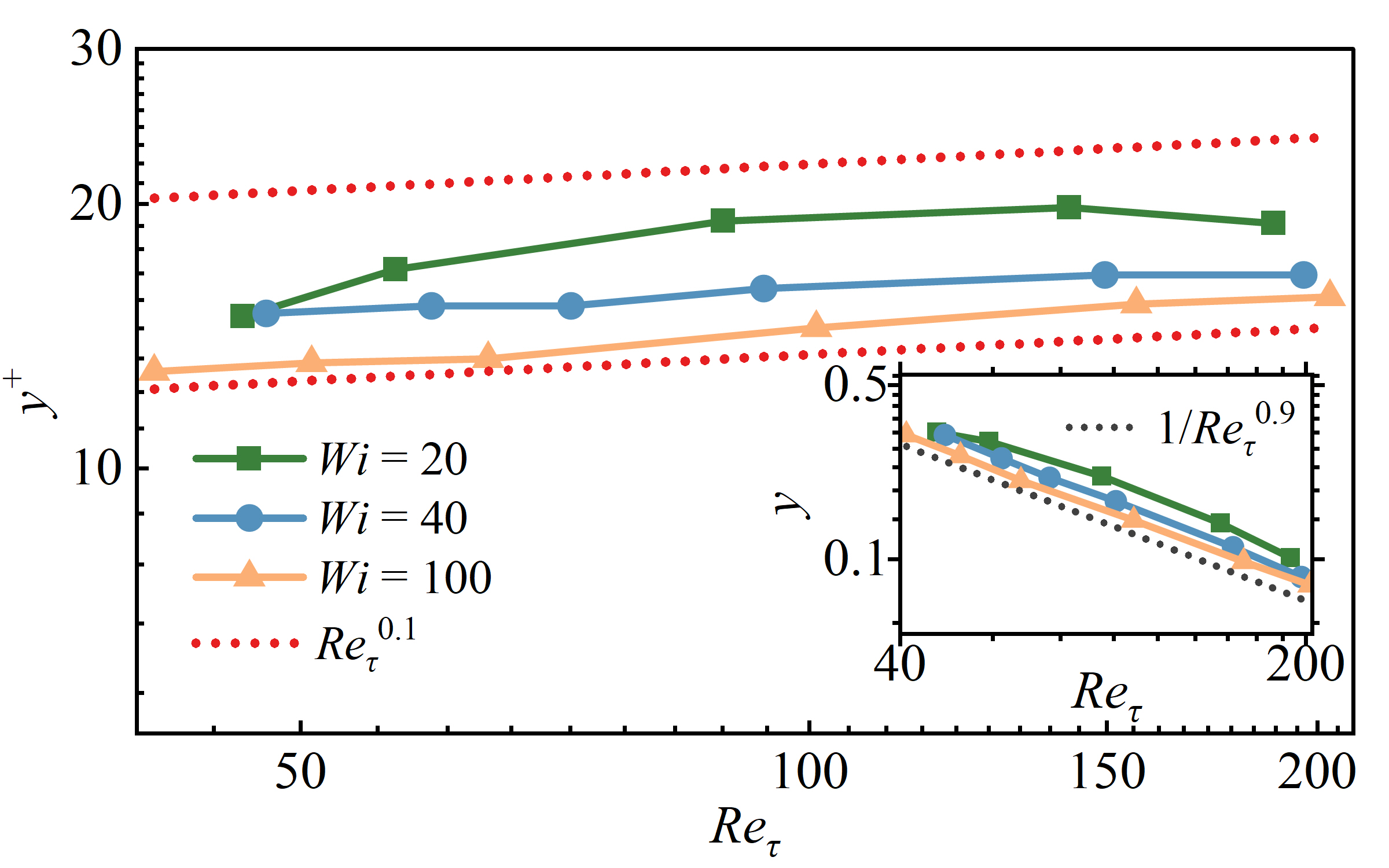}
	\caption{\label{fig10} Relationship between y+ (peak position of -G) and $Re_\tau$.}
\end{figure}

Figure \ref{fig10} displays the quantitative relationship between the peak location of $-G/C_f$ and $Re_\tau$ across different $Wi$, yielding a relatively weak Re dependence ($y^+_{-G_{\mathrm{max}}} \propto Re_\tau^{0.1}$). This 0.1 power law indicates that inertia can hardly alter the physical fact that polymer extension and energy exchange rely heavily on the strong near-wall shear. Consequently, the active energy-exchange region remains confined to the near-wall area adjacent to the viscous sublayer ($y^+ \approx 10 \sim 20$). We define this location as a fixed energy exchange critical layer (EECL), where a dynamic mechanism seems to exist that is almost unaffected by the force of inertia. We will focus on this discussion in the next section.

\begin{figure}
	\centering
	\includegraphics[width=1\textwidth]{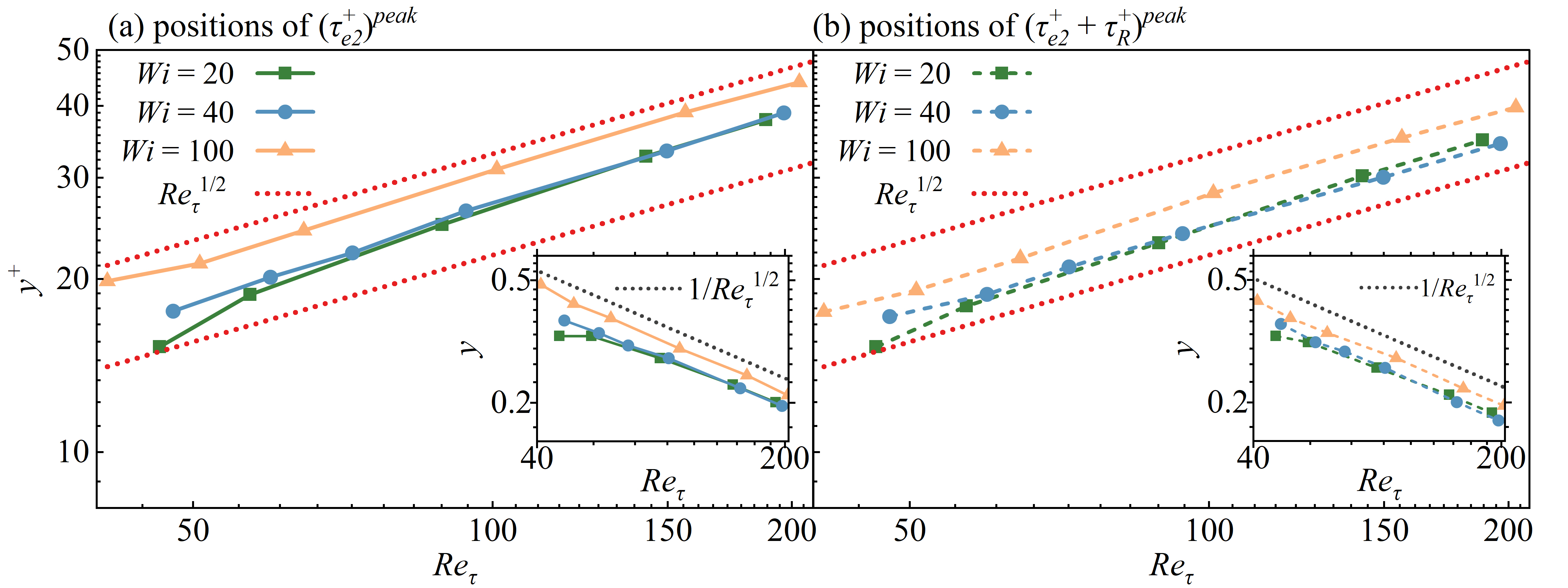}
	\caption{\label{fig11} Relationship between peak position $y^+$ of $\tau_e$ (and $\tau_e$+$\tau_R$) and $Re_\tau$.}
\end{figure}

Figure \ref{fig11}(a) displays the variation of the peak location of nonlinear elastic shear stress ($y^+_{\tau_e}$) with respect to the friction Reynolds number $Re_\tau$ under different $Wi$. It is evident that the peak location does not remain anchored in a fixed near-wall inner-scaled region. Instead, as inertia ($Re_\tau$) increases, $y^+_{\tau_e}$ exhibits a pronounced outward migration, strictly obeying a scaling law of $y^+_{\tau_e} \propto Re_\tau^{1/2}$. This $1/2$ power law is not an empirical data fit, but fundamentally originates from the intrinsic mean momentum balance of the 2D EIT. In classical Newtonian wall turbulence, the peak of the Reynolds shear stress is known to follow an identical $Re_\tau^{1/2}$ scaling, which signifies the location of the mesolayer (Afzal 1982; Lee and Moser 2015). As elucidated by Lee and Moser (2015), this location mathematically emerges from the total stress balance: the peak of the Reynolds stress occurs precisely where the rapid wall-normal decay of the viscous stress is perfectly balanced by the linear decrease of the total stress across the channel. Strikingly, based on the total shear stress equation for viscoelastic channel flow, we theoretically demonstrate that a highly analogous mechanism governs the 2D EIT. In our flows, since the mean velocity profile maintains a log-like region (figure \ref{fig6}), the viscous stress undergoes a similar sharp decay. Crucially, the Reynolds stress in 2D EIT is virtually completely suppressed ($\tau_R \approx 0$). Consequently, the nonlinear elastic shear stress must step in to fill the momentum deficit. By taking the derivative of the total stress balance, we find that the elastic stress must reach its global maximum at $y^+_{\tau_e} = f(Wi) \cdot Re_\tau^{1/2}$ (the detailed derivation is provided below). 

It should be noted that although $\tau_R$ is extremely small in 2D EIT, its negative nature is a distinctive feature of this flow state. To eliminate potential theoretical errors caused by completely neglecting $\tau_R$, we further track the evolution of the combined stress peak. The results in figure \ref{fig10}(b) confirm that the peak location of the combined stress, $y^+_{\tau_e+\tau_R}$, still robustly adheres to the $Re_\tau^{1/2}$ scaling law.
The remarkable consistency between our theoretical derivation, the DNS data, and the Newtonian mesolayer framework implies that this $1/2$ power-law peak location embodies profound underlying dynamics. Below this peak, the viscous stress dominates but decays rapidly, forcing the polymers to be strongly stretched to compensate for the decreasing total stress. Above this peak, the elastic shear stress becomes the dominant mechanism supporting the turbulent drag in the outer region. Therefore, $y^+_{\tau_e}$ marks the critical coordinate where a fundamental handover of the momentum transfer mechanism occurs. Based on this, we define this peak location as the Elasto-Inertial Critical Layer (EICL). In 2D EIT, the elastic shear stress flawlessly replicates the dynamical role of the Reynolds stress in traditional Newtonian turbulence; upon reaching the EICL, elasticity takes over the momentum transfer from viscosity. This $1/2$ scaling robustly proves the physical homology between the EICL in EIT and the mesolayer in Newtonian turbulence. In our previous work, we also identified the dominant role of elastic stress in the self-sustaining cycle (Zhang et al., 2021). Meanwhile, the influence of elasticity ($Wi$) on this position cannot be ignored, but its complexity temporarily hinders further theoretical derivation, preventing us from obtaining an explicit expression for the prefactor $f(Wi)$.

   \subsection{Self-similarity of Fluctuations and Self-Sustaining Mechanism}

The preceding analysis clearly demonstrates that as inertial effects strengthen, 2D EIT undergoes not only a dramatic surge in fluctuation intensity but also a significant spatial migration of its core dynamic layer (EICL) following specific $Re$-scaling laws. Such substantial evolution in macro-spatial structure and energy magnitude naturally raises a fundamental question: does the continuous enhancement of inertia alter the physical essence of EIT, or does it merely modulate the working space and amplitude of a universal self-sustaining mechanism? To address this, we examine the statistical features of instantaneous fluctuations through probability density functions (PDFs) at specific flow locations.Prior to this, we hypothesize that despite the strong Re dependence of 2D EIT, the core dynamical mechanism governing turbulence generation and maintenance remains universal and invariant. To test this, we analyze the statistical characteristics and transient topologies across various cases. 
Figure \ref{fig11} displays the PDFs of streamwise velocity ($u'$), wall-normal velocity ($v'$), and elastic shear stress ($\tau_e'$) at different wall-normal positions for a representative case ($Re=2000, Wi=40$). The distinct variation of PDFs with $y$ necessitates a careful selection of the statistical anchor. Given that the scaling exponent of the $-G/C_f$ peak location exhibits a much weaker dependence on inertia compared to the $\tau_e$ peak, the former is selected as the optimal location for extracting the underlying self-similarity of EIT.	

\begin{figure}
	\centering
	\includegraphics[width=0.8\textwidth]{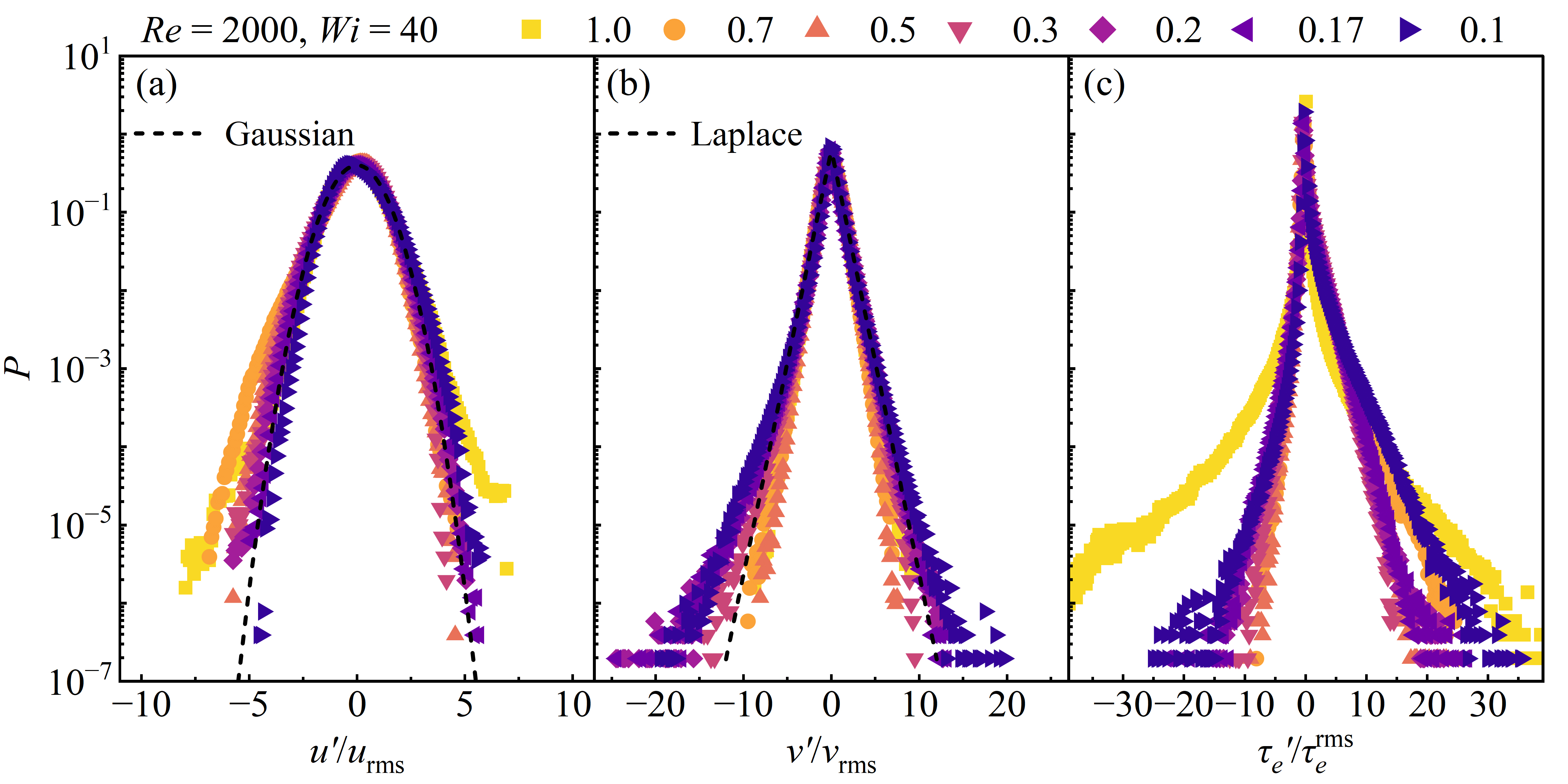}
	\caption{\label{fig12} Probability density functions (PDFs) of normalized fluctuations at different locations for $Wi = 40, Re = 2000$ across different Reynolds numbers. (a) Streamwise velocity fluctuation $u'/u_{\mathrm{rms}}$; (b) Wall-normal velocity fluctuation $v'/v_{\mathrm{rms}}$; (c) Elastic shear stress fluctuation $\tau_e'/\tau_{e,\mathrm{rms}}$.}
\end{figure}

\begin{figure}
	\centering
	\includegraphics[width=0.8\textwidth]{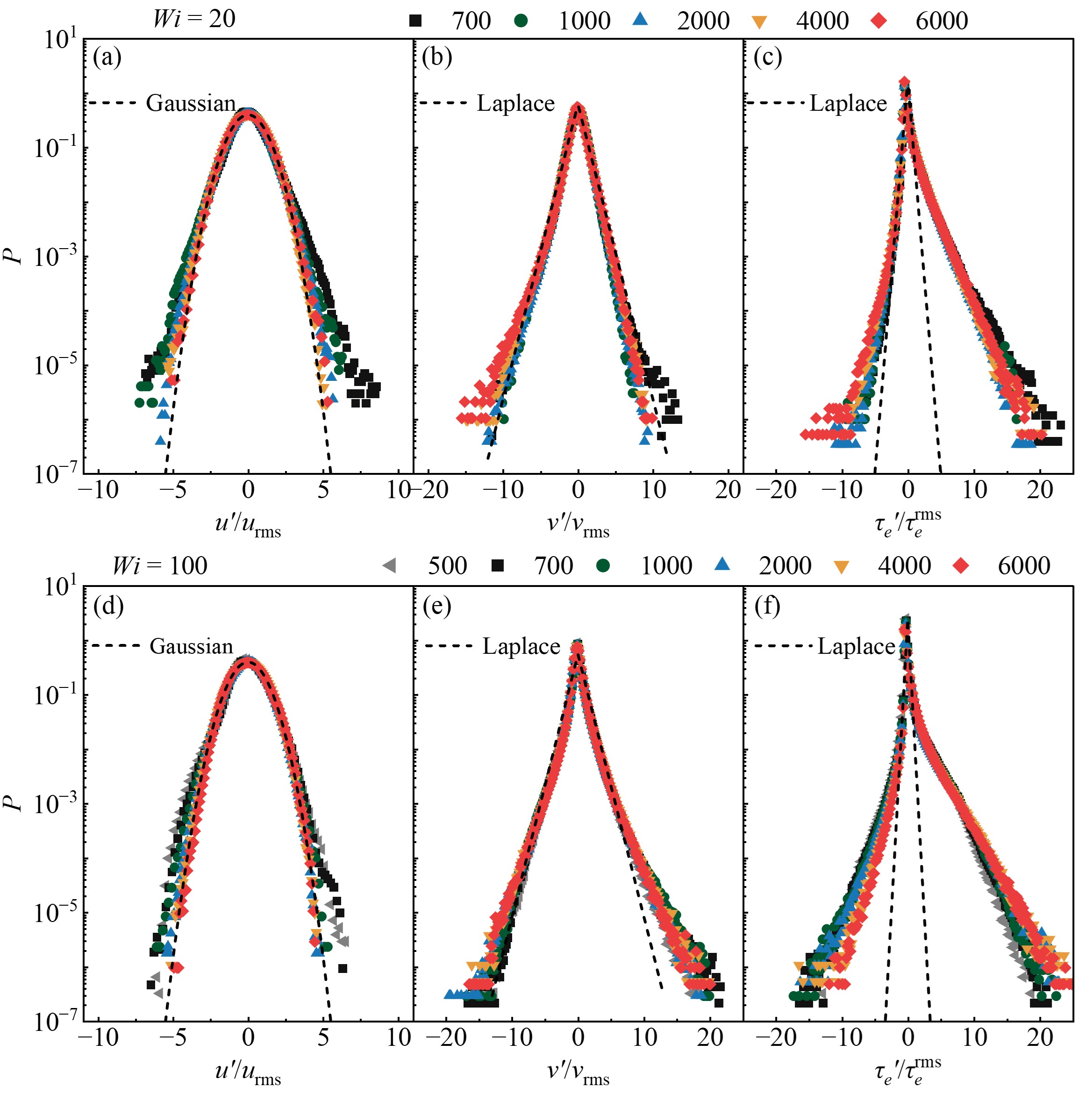}
	\caption{\label{fig13} Probability density functions (PDFs) of normalized fluctuations at the peak location of $-G/C_f$ for $Wi = 20, 100$ across different Reynolds numbers. (a, d) Streamwise velocity fluctuation $u'/u_{\mathrm{rms}}$; (b, e) Wall-normal velocity fluctuation $v'/v_{\mathrm{rms}}$; (c, f) Elastic shear stress fluctuation $\tau_e'/\tau_{e,\mathrm{rms}}$.}
\end{figure}

\begin{figure}
	\centering
	\includegraphics[width=0.6\textwidth]{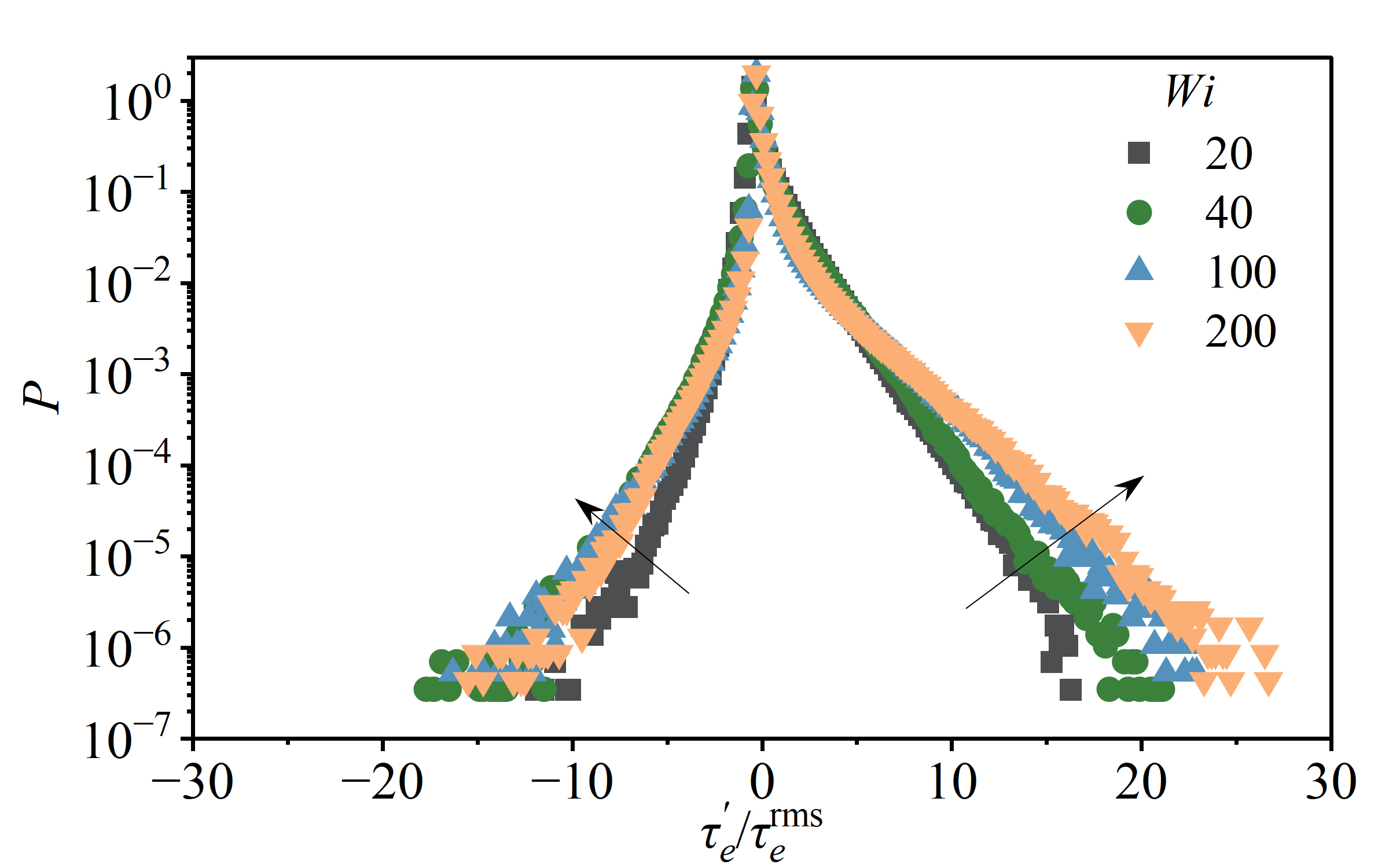}
	\caption{\label{fig14} PDFs of normalized elastic shear stress fluctuations for $Re$ = 2000 across different Wi.}
\end{figure}

Figure \ref{fig13} illustrates the PDFs of normalized fluctuations ($u'/u_{\mathrm{rms}}$, $v'/v_{\mathrm{rms}}$, and $\tau_e'/\tau_{e,\mathrm{rms}}$) extracted at the $-G/C_f$ peak across a range of $Re$ ($500 \sim 6000$) for fixed $Wi=20, 40,$ and $100$. Remarkably, all PDF curves exhibit an excellent collapse across the investigated $Re$ interval. This provides compelling evidence for a universal dynamical self-similarity within the core energy-exchange layer, independent of inertial strength. While similar asymptotic behaviors have been reported in elastic turbulence (ET) (Burghelea et al. 2006; Li et al. 2024), the present results show that in EIT, this collapse is established from the onset of the turbulent state along the inertial dimension. Comparison with Figure \ref{fig14} confirms that while inertia triggers the initial instability and sets the energy magnitude, the relative probability distribution of fluctuations is governed by elastic nonlinearity. This reinforces the dominant role of elastic effects in the dynamics of 2D EIT.	

\begin{figure}
	\centering
	\includegraphics[width=0.6\textwidth]{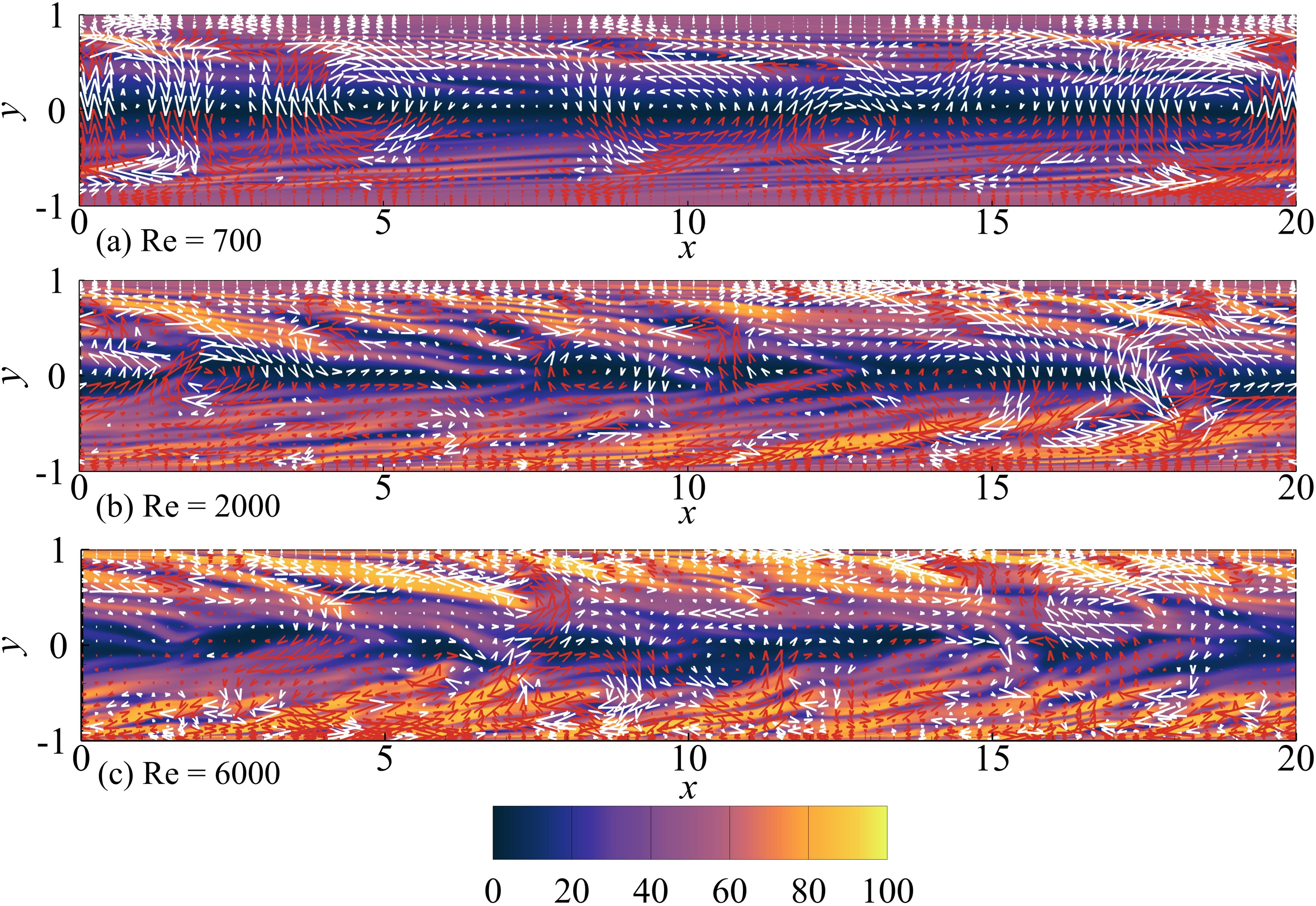}
	\caption{\label{fig15} Instantaneous field superposition of velocity vector, polymer extension and elastic shear stress fluctuations for Wi = 20.}
\end{figure}

\begin{figure}
	\centering
	\includegraphics[width=1\textwidth]{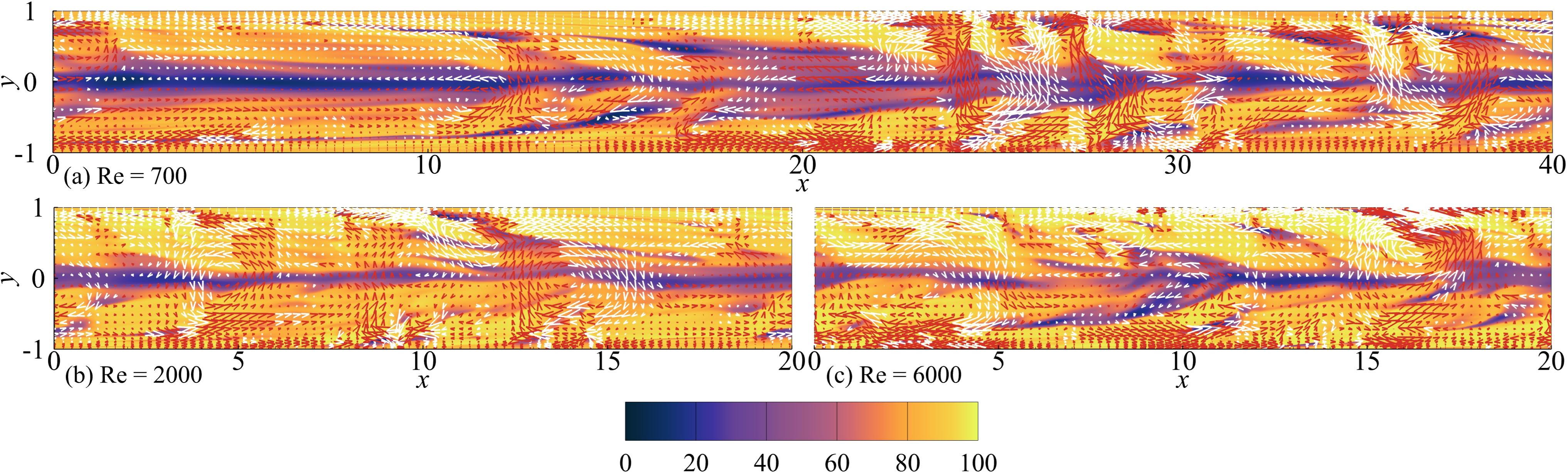}
	\caption{\label{fig16} Instantaneous field superposition of velocity vector, polymer extension and elastic shear stress fluctuations for Wi = 100.}
\end{figure}

Further inspection of the PDF topologies in figure \ref{fig13} reveals the physical origin of this universal mechanism. The distribution of $u'$ closely follows a Gaussian profile, indicating a passive response to the large-scale background shear with negligible intermittency. Conversely, $v'$ exhibits a skewed, heavy-tailed distribution, while $\tau_e'$ shows an exponential-like profile with extreme kurtosis. These significant deviations from Gaussianity signify the presence of intense intermittency and extreme events driven by elastic forces. We attribute these events to the structural evolution of polymer sheets, specifically their stretching, relaxation, and fracture. The synchronization between the negative tails of $v'$ and $\tau_e'$ suggests a causal link between extreme wall-normal motions and localized bursts of elastic stress.	
To substantiate this statistical inference, we analyze the spatial coupling between the instantaneous polymer extension, velocity vectors, and elastic stress fluctuations (figures \ref{fig15} and \ref{fig16}). Quadrant analysis reveals that regions of negative elastic fluctuations ($\tau_e' < 0$) align precisely with "weak extension zones" where polymer sheets undergo local relaxation or fracture. Crucially, these regions are dominated by third-quadrant (Q3) events ($u' < 0, v' < 0$). This suggests that low-speed fluid elements impacting toward the wall impose transverse shear and impact on the polymer extension structures, leading to relaxation or even fracture. Consequently, local elastic shear stress (negative value region) decreases significantly below the average value.	

Combining the statistical evidence from figure \ref{fig13} and the transient topology in figures \ref{fig15} and \ref{fig16}, we construct a closed-loop self-sustaining mechanism for 2D EIT: polymers accumulate elastic energy through continuous stretching by the near-wall shear and Q1 events; subsequently, occasional and discrete Q3 events trigger localized relaxation or fracture of high-stretched sheet structures, leading to the release of stored elastic energy (corresponding to the extreme negative tail in $\tau_e'$ PDF). This released energy is efficiently converted into the TKE of surrounding fluid, providing a continuous energy supply to the flow. This spatiotemporal dynamic cycle constitutes the dynamics enabling EIT to achieve turbulent self-sustaining. While inertia performs spatial modulation and amplitude regulation by altering the background shear environment, the elastic nonlinearity remains the primary driver of the universal statistical essence of EIT.

Although an $Re$-independent statistical self-similarity is observed in terms of local statistical properties, this by no means implies that inertial effects in EIT can be neglected. Furthermore, it does not justify simply equating the core dynamical mechanism of EIT to that of ET. The coexistence of this statistical self-similarity with the aforementioned inertial modulations on the statistical characteristics and flow structures reveals a synergistic yet decoupled relationship between elastic and inertial nonlinearities within EIT. Specifically, the dominant elastic nonlinearity is primarily responsible for dictating the universal, underlying dynamical self-sustaining mechanism of EIT. Conversely, by altering the near-wall shear and convective environment, the inertial effect predominantly acts to regulate the amplitudes of statistical quantities and facilitate the wall-ward migration of core flow structures.
	
	\section {Concluding remarks}\label{4}

By performing DNSs of two-dimensional viscoelastic channel flow driven by a constant flow rate, this paper systematically investigates the effects of inertia on the quantitative characteristics and physical processes of EIT. For the first time, a $1/2$-power scaling law between the peak position of elastic shear stress and $Re_\tau$ in 2D EIT is established, the concept of EICL is proposed, and the universal self-sustaining mechanism of EIT under inertial modulation is revealed. These findings provide a quantitative basis for understanding the role of inertia in EIT. The main conclusions are summarized as follows:
    (1) Enhanced inertia promotes the breakdown of vortex structures from a large-scale sparse distribution into high-density clusters of micro-vortices, forcing them to shift from the channel center and closely attach to the wall. Meanwhile, the amplitude of polymer stretch, the contribution of nonlinear elastic shear stress, and the turbulent kinetic energy are significantly enhanced. The root-mean-square of streamwise velocity fluctuations ($u_{\mathrm{rms}}$) exhibits a bimodal characteristic, which is associated with the head-to-tail gradients of the sheet-like polymer stretch structures. The evolution of the wall-normal velocity fluctuations ($v_{\mathrm{rms}}$) with $Re$ is regulated by the relative strength between elasticity and inertia.
    (2) The peak position of energy conversion ($y^+_{-G_{\mathrm{max}}}$) migrates with the friction Reynolds number ($Re_\tau$) following a $0.1$-power law, indicating that the energy-active region remains strictly confined near the viscous sublayer. In contrast, the peak position of the nonlinear elastic shear stress ($y^+_{\tau_e}$) strictly obeys a scaling law of $y^+_{\tau_e} \propto Re_\tau^{1/2}$. Theoretical derivations prove that this scaling originates from the mean momentum balance. We define $y^+_{\tau_e}$ as the elasto-inertial critical layer (EICL), marking the exact transition point where viscous stress decays and elastic shear stress takes over the dominant role in momentum transport. Unlike in Newtonian fluids where the Reynolds stress peak is anchored at a fixed inner scale, the EICL migrates significantly with inertia.
    (3) At the energy conversion peak, the PDFs of the streamwise velocity, wall-normal velocity, and elastic shear stress fluctuations collapse remarkably across different $Re$. This proves that the underlying turbulence generation mechanism of EIT possesses cross-Reynolds-number self-similarity, remaining fundamentally unchanged despite the enhancement of inertia. The PDF of streamwise velocity fluctuations approximates a Gaussian distribution, while that of wall-normal velocity fluctuations exhibits a Laplacian distribution. The elastic shear stress fluctuations follow an exponential-like distribution with a positively skewed fat tail, indicating a strong correlation between extreme wall-normal motions and the intense bursts of elastic stress. Instantaneous topological analysis reveals that the third-quadrant (Q3) sweep events exert transverse impacts on the sheet-like polymer stretch structures, causing local rupture and relaxation. This process releases elastic energy and converts it into turbulent kinetic energy, thereby forming a self-sustaining closed loop: Q1 forward stretching $\rightarrow$ Q3 reverse impact $\rightarrow$ stretch structure rupture and relaxation $\rightarrow$ rapid release of elastic energy $\rightarrow$ explosive regeneration of turbulent kinetic energy.
In summary, future work will be dedicated to extending these scaling laws and self-sustaining mechanisms to a broader parameter space ($\beta$, $Wi$, $Re$) and to three-dimensional EIT, as well as conducting systematic research on purely elastic turbulence (ET) without inertial effects. Based on these efforts, a more precise quantitative correlation between ET and EIT will be established.

\section*{Theoretical derivation of the peak location of elastic shear stress in 2D EIT}	
	Based on the total mean momentum balance for a two-dimensional channel flow (assuming the Reynolds shear stress is negligible), the nonlinear elastic shear stress $\tau_e$ can be expressed as:
	\begin{equation}
		\tau_e = \tau_w \left(1 - \frac{y}{H}\right) - \nu \frac{\mathrm{d}\bar{u}}{\mathrm{d}y}
	\end{equation}
	where $\tau_w$ is the wall shear stress, $H$ is the channel half-height, $\nu$ is the kinematic viscosity, and $\bar{u}$ is the mean streamwise velocity.
	To locate the global maximum of the elastic shear stress, we take the derivative of $\tau_e$ with respect to the wall-normal coordinate $y$ and set it to zero:
	\begin{equation}
		\frac{\mathrm{d}\tau_e}{\mathrm{d}y} = 0
	\end{equation}
	Substituting equation (1) into equation (2) yields:
	\begin{equation}
		-\frac{\tau_w}{H} - \nu \frac{\mathrm{d}^2\bar{u}}{\mathrm{d}y^2} = 0 \quad \Rightarrow \quad \frac{\tau_w}{H} = -\nu \frac{\mathrm{d}^2\bar{u}}{\mathrm{d}y^2}
	\end{equation}
	Next, the functional form of the mean velocity gradient $\mathrm{d}\bar{u}/\mathrm{d}y$ must be determined. We hypothesize that $\partial\bar{u}/\partial y = f(u_\tau, y, \nu)$. Using dimensional analysis (the Buckingham Pi theorem), we construct the first dimensionless group:
	\begin{equation}
		\Pi_1 = \frac{\partial \bar{u}}{\partial y} u_\tau^a y^b \sim (T^{-1})(L T^{-1})^a L^b = L^{a+b} T^{-1-a}
	\end{equation}
	Equating the exponents to zero gives $a = -1$ and $b = 1$, yielding $\Pi_1 = \frac{y}{u_\tau} \frac{\partial \bar{u}}{\partial y}$. Similarly, constructing a second dimensionless group $\Pi_2 = \nu u_\tau^c y^d$ yields $c = -1$ and $d = -1$, resulting in $\Pi_2 = \frac{\nu}{u_\tau y} = \frac{1}{y^+}$. We can redefine this group as $\Pi_2' = y^+$. Based on the functional relationship between these dimensionless groups $\Pi_1 = f(\Pi_2')$, we obtain:
	\begin{equation}
		\frac{\mathrm{d}\bar{u}}{\mathrm{d}y} = \frac{u_\tau}{y} f(y^+)
	\end{equation}
	In the log-like region of the mean velocity profile ($y^+ \gg 1$), the direct viscous effects are weak, and the parameter $\nu$ should not explicitly appear. Therefore, $f(y^+)$ must approach a constant $C$. In this regime, the velocity gradient and its second derivative become:
	\begin{equation}
		\frac{\mathrm{d}\bar{u}}{\mathrm{d}y} = C \frac{u_\tau}{y}
	\end{equation}
	\begin{equation}
		\frac{\mathrm{d}^2\bar{u}}{\mathrm{d}y^2} = -C \frac{u_\tau}{y^2}
	\end{equation}
	Let $y_{max}$ denote the peak location of the elastic shear stress. Substituting equation (7) back into the derivative balance equation (3), we get:
	\begin{equation}
		\frac{\tau_w}{H} = \nu C \frac{u_\tau}{y_{max}^2}
	\end{equation}
	Using the kinematic definition of the wall shear stress, $\tau_w = u_\tau^2$, we can solve for $y_{max}$:
	\begin{equation}
		y_{max} = \sqrt{\frac{\nu C u_\tau H}{\tau_w}} = \sqrt{\frac{\nu C H}{u_\tau}}
	\end{equation}
	Finally, converting this physical location into the inner-scaled dimensionless coordinate $y_{max}^+$:
	\begin{equation}
		y_{max}^+ = \frac{y_{max} u_\tau}{\nu} = \sqrt{\frac{C H u_\tau}{\nu}} = \sqrt{C \cdot Re_\tau} = f(Wi)\cdot\sqrt{Re_\tau}
	\end{equation}
	Extracting the proportionality, we mathematically demonstrate that the peak location of the elastic shear stress strictly follows the scaling law:
	\begin{equation}
		y_{max}^+ \propto Re_\tau^{1/2}
	\end{equation}

	\section*{Acknowledgements}
	This research was funded by the National Natural Science Foundation of China (NSFC 52006249, 12202308, 12472255).
	
	\section*{Declaration of interests}
	The authors report no conflict of interest.
	
	%\bibliographystyle{jfm}
	%\bibliography{jfm}
	%Use of the above commands will create a bibliography using the .bib file. Shown below is a bibliography built from individual items.
	
	\bibliographystyle{jfm}
	%\bibliography{jfm2esam}

	%% End of file `jfm2esam.bib'.
	
\end{document}